\documentclass[manuscript]{acmart}
\AtBeginDocument{%
  }

\setcopyright{acmlicensed}
\copyrightyear{2018}
\acmYear{2018}
\acmDOI{XXXXXXX.XXXXXXX}
\acmConference[Conference acronym 'XX]{Make sure to enter the correct
  conference title from your rights confirmation email}{June 03--05,
  2018}{Woodstock, NY}
\acmISBN{978-1-4503-XXXX-X/2018/06}

\usepackage{natbib}
\usepackage{booktabs}
\usepackage{longtable}
\usepackage{tabularx}
\usepackage{threeparttable}
\usepackage{ragged2e}
\usepackage{makecell}
\usepackage{array}
\newcolumntype{L}[1]{>{\RaggedRight\arraybackslash}p{#1}}
\newcolumntype{Y}{>{\RaggedRight\arraybackslash}X}
\usepackage{placeins}
\usepackage{enumitem}
\usepackage{float}
\usepackage[table]{xcolor}

\definecolor{sectiongray}{RGB}{242,242,242}

\setlist[itemize]{leftmargin=*, nosep}
\begin{document}

\title{"MeBo Leaves a Piece of You Behind": Designing a Relational Voice-Based Memory Companion for Older Adults}

\author{Hasibur Rahman}
\authornote{Both authors are co-first authors.}
\affiliation{%
  \institution{Northeastern University}
  \city{Boston}
  \state{Massachusetts}
  \country{USA}}
  \email{rahman.has@northeastern.edu}

\author{Mahsa Nasri}
\authornotemark[1]
\affiliation{%
  \institution{Northeastern University}
  \city{Boston}
  \state{Massachusetts}
  \country{USA}}
  \email{nasri.m@northeastern.edu}

\author{Manasi Vaidya}
\affiliation{%
  \institution{MIT Media Lab}
  \institution{Massachusetts Institute of Technology (MIT)}
  \city{Cambridge}
  \state{Massachusetts}
  \country{USA}}
  \email{manasiv@mit.edu}

\author{Melika Vafafar}
\affiliation{%
  \institution{Northeastern University London}
  \city{London}
  \country{United Kingdom}}
  \email{vafafar.m@northeastern.edu}

\author{Jessie Chin}
\affiliation{%
  \institution{University of Illinois Urbana-Champaign}
  \city{Urbana}
  \state{Illinois}
  \country{USA}}
  \email{chin5@illinois.edu}

\author{Smit Desai}
\authornote{Corresponding author}
\affiliation{
  \institution{Northeastern University}
  \city{Boston}
  \state{Massachusetts}
  \country{USA}}
\email{sm.desai@northeastern.edu}

\renewcommand{\shorttitle}{MeBo Leaves a Piece of You Behind}
\renewcommand{\shortauthors}{Rahman et al.}

\begin{abstract}
Autobiographical remembering supports identity, well-being, and social connection in later life, yet voice-based memory technologies largely rely on isolated prompts. We designed and built MeBo, a fully functional relational voice-based memory companion, through participatory design with 11 older adults. Their accounts shaped four Design Strategies that guided MeBo’s interaction design and multi-agent implementation. In a mixed-methods evaluation with 20 older adults, participants found MeBo exceptionally usable (SUS = 87.75), enjoyable, sociable, emotionally responsive, and trustworthy. Participants reported higher positive affect and momentary social connection and lower negative affect after the session than before. Participants described how MeBo followed their stories, returned to earlier memories, adapted to their preferences, and made its growing memory visible and controllable. MeBo’s relational framing surfaces tensions around what it should remember, who may access memories produced through interaction, and what becomes of them when the user or MeBo is no longer present.

\end{abstract}

\ccsdesc[500]{Human-centered computing~Empirical studies in HCI}
\ccsdesc[300]{Human-centered computing~Natural language interfaces}

\keywords{Older adults; conversational AI; voice user interfaces; reminiscence; relational agents; autobiographical memory; social connectedness; human–AI interaction.}



\begin{teaserfigure}
  \centering
  \includegraphics[width=\textwidth]{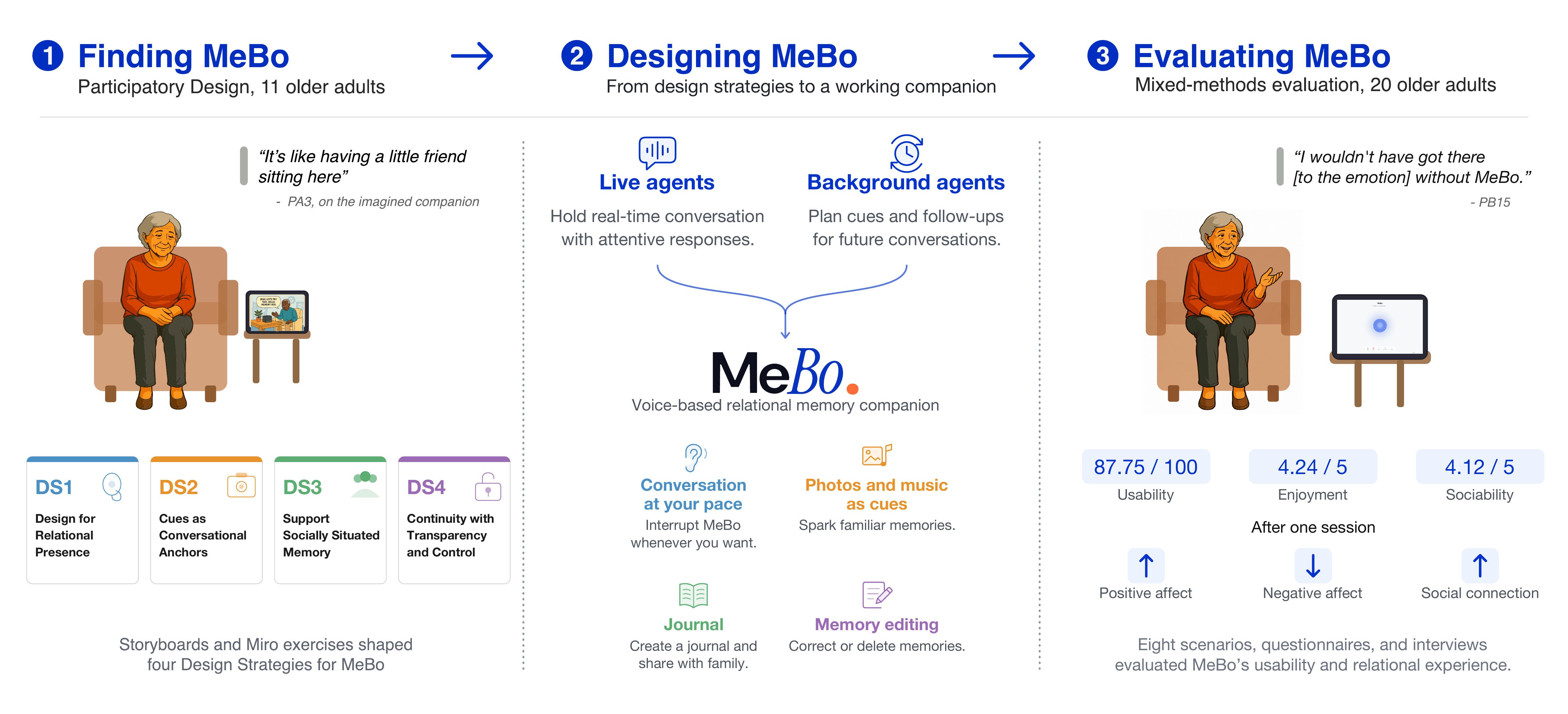}
  \caption{Overview of MeBo: participatory design with 11 older adults yields four Design Strategies (Finding MeBo), which shape a multi-agent voice companion (Designing MeBo), evaluated with 20 older adults across usability, relational, and affective outcomes (Evaluating MeBo).}
  \Description{Three panels connected by left-to-right arrows trace Finding, Designing, and Evaluating MeBo. Finding MeBo shows an older adult beside a storyboard and the quote, “It’s like having a little friend sitting here,” attributed to PA3. Participatory design with 11 older adults produces four strategies: relational presence, cues as conversational anchors, socially situated memory, and continuity with transparency and control. Designing MeBo connects live agents for attentive conversation and background agents for future cues and follow-ups to the working companion. Four features illustrate this translation: conversation at the user's pace with interruption, photographs and music as cues, a journal that can be shared with family, and memory correction or deletion. Evaluating MeBo shows an older adult speaking beside the voice interface and reports results from 20 participants: usability 87.75 out of 100, enjoyment 4.24 out of 5, and sociability 4.12 out of 5. Arrows indicate increased positive affect, decreased negative affect, and increased social connection after one session. PB15 states, “I wouldn't have got there [to the emotion] without MeBo.” Eight scenarios, questionnaires, and interviews connect the system to these findings.}
  \label{fig:teaser}
\end{teaserfigure}

\maketitle

\section{Introduction}
\begin{quote}
\emph{``I really felt like I was back in the moment, back in time, back in those days. I was seeing where I used to play baseball, where we used to go for bicycle rides, where we used to hang out at movie theaters and at the White Castle---all the places I had talked about. It is almost like having a dream, but being awake while you do it.''}

\hfill---PB4, on reminiscing with MeBo
\end{quote}

\noindent PB4 describes an experience of remembering that extends beyond simply retrieving information about the past. Through conversation, memories became vivid, situated, and open to reflection. Yet prior technologies for older adults have often treated memory as something to be prompted, recorded, or supported through discrete interventions \cite{zhang2026memorymeaning,lazar_systematic_2014,pardini2025exploring,moon2020effect}. This leaves less room for conversational systems to learn personal memories over time, revisit them when relevant, and support the relationships that sustain their meaning. In this paper, we explore this broader possibility through MeBo, a relational voice-based memory companion for older adults.

Designing for remembering in later life is increasingly consequential as the older-adult population grows. By 2030, one in six people worldwide will be 60 years or older; by 2050, this population is projected to reach 2.1 billion\footnote{\url{https://www.who.int/news-room/fact-sheets/detail/ageing-and-health}}. Autobiographical memories are recollections of one's own life that people revisit and make meaning from, such as childhood games, a wedding, or family holidays. In later life, these memories sustain conversation, identity, intimacy, and emotional regulation \cite{webster1993functions,bluck1998reminiscence,westerhof2014celebrating}. Photographs, music, places, voices, and other cues create an occasion for such memories to surface and acquire new meaning in the present \cite{zhang2026memorymeaning,otenen2024multimodal,gayler2020flavor,piper2013audio}. Remembering extends across relationships; family members prompt one another, fill in details, and pass stories between generations \cite{jones2018coconstructing,liaqat2022hint,li2023arstorytelling,lindley2012before}. Narrators deliberately keep some memories untold, preserving authority over when and with whom they are shared \cite{jones2021mysteries}. These practices make autobiographical memory a personal and social activity, letting older adults connect to the past and continually shape memories in the present. Systems supporting remembering must therefore account for the personal histories, emotions, and relationships that give memories meaning.


Because remembering is often accomplished through conversation, voice offers a particularly relevant modality for technologies seeking to participate in these practices. Voice user interfaces (VUIs)\footnote{A VUI is the speech-based medium through which a person interacts with a conversational agent.} rely on speech for interaction, reducing demands associated with typing, touchscreens, mouse use, and sustained visual attention \cite{desai2023storytelling,Pradhan2018,Corbett2016}. Older adults with little experience, therefore, effectively engage with VUIs, although recognition, turn-taking, pacing, repair, and routine fit continue to shape accessibility in practice \cite{pradhan2020intelligent,huang_designing_2025,wolters2009old,chen2021barriers}. Older adults also interpret VUIs through instrumental and social frames, shaping expectations about conversational style, attention, familiarity, and appropriate conduct \cite{pradhan2019,chin2024aunt,hu2022polite}.

HCI research has explored how digital systems can help older adults create and share memory artifacts with family, and how they can prompt and elaborate autobiographical memories through photographs, audio, location histories, and generated music \cite{zhang2026memorymeaning,Zhang,piper2013audio,white2023memorytracer,jin2024generative,waycott2013producers,li2022memento}. Recent conversational systems extend through dialogue, multimodal cues, mutual reminiscence, and access to prior conversational context \cite{sun_chorus_2025,jiang2025remini,zulfikar2024memoro}. These studies suggest that conversational support can help people elaborate and connect through memories, while requiring care in handling emotional complexity, protecting social information in personal images, and pacing the reappearance of personal data over time \cite{hsu_bittersweet_2025,kandappu2021privacyprimer,odom2019olly}.

Research on relational agents suggests that continuity supports familiarity and engagement when agents maintain conversational coherence, variety, and appropriate responsiveness \cite{bickmore2005relationship,bickmore2005acceptance,bickmore2010maintaining,vardoulakis2012companions}. Persistent memory can further strengthen believability, engagement, self-disclosure, social identity, and trust \cite{richards2014forgetmenot,cox2023previous,jo2024carecall,jiang2026recallbot}. However, incorrect recall, verbatim references, and memories of sensitive topics can undermine believability and raise privacy concerns \cite{richards2014forgetmenot,cox2023previous,jo2024carecall}. In home settings, retained information also raises questions of ownership, access, correction, and disclosure \cite{luria2020boundaries,singh2026caregivers,karimi2025collaborative}.

A voice-based memory companion sits at the intersection of these research areas, requiring accessible conversation, continuity across memories, relational interaction, and user authority. What we lack is an end-to-end account: how older adults' participatory contributions can shape such a companion, how they survive implementation, and how older adults experience the resulting system in use. 


To address this gap, we present MeBo, a relational voice-based memory companion for older adults, developed through participatory design, implemented as a working system, and evaluated in use. MeBo moves beyond an intervention organized around isolated prompts to a voice-based relational companion that carries memories across conversations, returns to them over time, and makes that continuity visible and controllable. MeBo began as Memory Box, a participatory design concept for a voice-based reminiscence intervention that prompts, gathers, and preserves memories. Participants in the participatory study shifted the center of design from a bounded reminiscence tool to a companion that could become familiar, remain available as memories surfaced in everyday life, and support relationships with family and friends through sharing and retelling memories. \emph{Memory Box} became MeBo.

Building on \citet{bickmore2005relationship}'s definition of relational agents, we use \emph{relational memory companion} to describe this form of participation in remembering. More specifically, we are referring to a memory companion designed to establish and maintain a long-term social-emotional relationship through attentive conversation and user-governed autobiographical continuity across encounters, including memories made and shared with others.

We conducted a mixed-methods usability study with 20 older adults. Participants found MeBo highly usable (SUS $87.75$), experienced it as relational, and described remembering with it as nostalgic and uplifting. After one session, positive affect and momentary social connection increased, while negative affect decreased. However, the continuity that made MeBo familiar also raised concerns about what it retained and what would become of it over time. We have three primary contributions:

\begin{itemize}
    \item Four Design Strategies, developed with 11 older adults, for designing relational voice-based memory companions that support relational presence, anchor conversation in personal memories, situate remembering within social relationships, and make memory continuity transparent and controllable;

    \item MeBo, a working voice-based memory companion that instantiates these strategies through persistent memory, multimodal cues, and user-governed family participation; and
    
    \item A mixed-method evaluation with 20 older adults showing how usability, relational experience, memory control, adoption boundaries, and immediate emotional and social effects shape relational memory companionship.
\end{itemize}

\section{Related Work}

We review three literatures that inform the design of a memory companion: how older adults encounter voice as both an interaction medium and a social presence; how later-life reminiscence technologies frame remembering as a dedicated activity; and how relational agents sustain engagement and memory across encounters.

\subsection{Conversational AI and Voice Technologies for Older Adults}

Voice interaction can reduce reliance on typing and complex graphical navigation, easing demands associated with conventional input and sustained visual attention \cite{desai2023storytelling,Pradhan2018,Corbett2016}. Accessibility, however, requires more than speech \cite{pradhan2020intelligent,huang_designing_2025,wolters2009old}. Initial encounters with smart speakers may be exploratory \cite{pradhan2020intelligent,kim2021first}, but continued use depends on discovering capabilities, fitting the system into existing routines, and having a reason to return \cite{trajkova2020toy,upadhyay2023exploration,yu2023history}. Longitudinal studies document growing familiarity but divergent patterns of use---sustained, low, declining, and late-surging---shaped by habits and socioemotional support \cite{kim2021longitudinal,chen2026trajectories}. Community support also helps sustain adoption \cite{karkera2023community}. Approachability alone, therefore, does not predict whether a voice system becomes meaningful in everyday life \cite{kim2021longitudinal,upadhyay2023exploration,chen2026trajectories,karkera2023community}, a distinction we examine in evaluating MeBo (\S\ref{sec:evaluation}).

Conversational interaction creates demands around system state, turn-taking, pacing, recognition, and repair \cite{pradhan2020intelligent,huang_designing_2025,chen2021barriers,brewer2022health}. Variation in speech intensity, voice quality, and pausing can increase recognition errors for older adults \cite{cohn2026asr}, and such breakdowns can disproportionately shape acceptance during learning \cite{desai2023learn}. Older adults value usable interaction alongside credibility, compassion, and control \cite{desai2023storytelling,brewer2022health}, while voice-first aging-in-place deployments show how lived context produces divergent needs and interpretations \cite{cuadra2023voicefirst}. These findings favor observable listening, turn-state correction, and visual or touch complements over a single interaction profile \cite{huang_designing_2025,hu2025beyond,brewer2023equitable,cohn2026asr}.

These interactional qualities shape how voice is socially understood. Older adults interpret VUIs through both instrumental and social frames \cite{pradhan2019}, and longitudinal use includes personal questions, advice-seeking, and interactions during stress \cite{oewel2023longtermcare}. Social framing need not position VUIs as substitutes for human relationships \cite{pradhan2019,Joshi_Ulabhaje_Nataraj_Martin-Hammond_2025,wong2024mentalhealth}. Instead, conversational style shapes acceptance and the social roles attributed to a VUI: preferences span formal, informal, polite, and direct styles \cite{chin2024aunt,hu2022polite}, and participants reject patronizing interaction \cite{horstmann2023patronized}. Co-design studies similarly identify persona, voice, and personalization as components of VUI sociability \cite{desai2023experience,hu2025beyond,kramer2021eating}. These plural and revisable relational configurations directly inform MeBo’s participatory persona design (\S\ref{sec:pd}) and the evaluation of its voice, persona, enjoyment, and sociability (\S\ref{sec:evaluation}).

This relational view is increasingly visible in voice-based memory technologies. Recent systems combine flexible dialogue, multimodal cues, and retained personal context \cite{liu2025bargein,wang_promoting_2024,sun_chorus_2025}, while participatory studies in aging and dementia care connect conversational roles, emotional boundaries, and family involvement \cite{lima_role_2023,singh2026caregivers,karimi2025collaborative}. Memory assistance systems also provide speech-based functional support \cite{cofre_voice_2020}; however, companion-robot studies show that improved latency, recognition, and conversational naturalness do not necessarily strengthen companionship \cite{satake2026refinement}. Together, this work positions voice as a relational medium shaped by personalization, empathy, multimodality, and control beyond its role as an access modality \cite{huang_designing_2025,pradhan2019,desai2023experience,hu2025beyond}. Memory companionship brings this relational view into everyday autobiographical remembering and meaning-making.

\subsection{Remembering in Later Life}
In later life, autobiographical remembering supports conversation, identity, intimacy, and emotional regulation beyond clinical settings \cite{webster1993functions,bluck1998reminiscence,westerhof2014celebrating}. It also occurs amid retirement, bereavement, changes in health and mobility, and social disconnectedness, which reshape social networks and opportunities for interaction \cite{nasem2020,courtin2017,cornwell2009}. Everyday remembering involves noticing a memory, deciding whether to pursue or elaborate it, and sometimes involving others \cite{blok2021facebook,cosley2012everyday,peesapati2010pensieve}; it is woven into ongoing life.

Reminiscence therapy structures this broader practice through facilitated sessions and media prompts and has been associated with improvements in psychological well-being, life satisfaction, social connection, and depressive symptoms among older adults \cite{lazar_systematic_2014,yen_systematic_2018}. Technology-mediated approaches deliver these interventions through digital media \cite{lazar_systematic_2014,pardini2025exploring}. Moon et al. organized a tablet intervention as eight 30-minute sessions over four weeks, each comprising an introduction, reminiscence, and wrap-up \cite{moon2020effect}. HCI systems often target remembering as a dedicated intervention; evaluations can overlook broader aims such as identity and emotional support \cite{zhang2026memorymeaning}.

Across these interventions, memory cues provide openings for remembering. Older adults connect self-defining memories to layered emotional cues \cite{sas2018selfdefining}, including photographs, voices, locations, sound, and flavor \cite{otenen2024multimodal,gayler2020flavor,piper2013audio,white2023memorytracer}. Systems also support digital content production \cite{thiry2013authoring,waycott2013producers} and intergenerational memento storytelling \cite{li2022memento}. ReminiBuddy uses artifact cues and multi-agent dialogue to elaborate older adults' memories \cite{sun_chorus_2025}; Remini guides reciprocal storytelling between loved ones \cite{jiang2025remini}. AI-generated music \cite{jin2024generative} and linked cross-generational photographs \cite{kang2021momentmeld} similarly propose associations without fixing their meaning. Media participate in remembering by prompting recall and supporting elaboration \cite{zhang2026memorymeaning,Zhang,waycott2013producers,li2022memento}. Remembering thus spans the person, artifacts, and, when stories are shared, other people, consistent with distributed and extended cognition \cite{hollan2000distributed,clark1998extended,sutton2010psychology}.

Remembering also circulates through relationships \cite{jones2018coconstructing,liaqat2022hint,li2023arstorytelling,lindley2012before}. Older adults and family members jointly discover, reconstruct, extend, and pass on stories across co-located and remote interactions \cite{jones2018coconstructing,liaqat2022hint,li2023arstorytelling,kang2021momentmeld}. This can become transactive as relatives prompt one another, supply missing details, and rely on who knows what \cite{wegner1991transactive,jones2018coconstructing,liaqat2022hint}. Social VR and digital storytelling support shared reflection and connection \cite{baker2021schoolsback,hausknecht2019wisdom}, while memory gifts, family histories, and legacy practices orient personal media toward present and future audiences \cite{gibson2023memorymachine,lindley2012before,thangaraj2026legacy}. Such circulation includes withholding: untold stories can protect relationships, identity, and narrative authority \cite{jones2021mysteries}. Commercial life-story applications, however, often favor privately owned legacy artifacts over ongoing relational storytelling \cite{schmidt2025lifestory}. Family memory is therefore negotiated participation over who may contribute, revise, or receive, rather than frictionless transfer \cite{jones2018coconstructing,jones2021mysteries,lindley2012before,thangaraj2026legacy}.

The limits of this intervention framing also emerge in a 2026 review by \citet{zhang2026memorymeaning}, who note that LLM-based reminiscence systems are ``frequently framed as single-session encounters, evaluated mainly through conversational fluidity or story quantity'' and call for ``longer-term, slower engagements'' supporting reflection, narrative evolution, and emotional processing across time. MeBo addresses these temporal and relational challenges by carrying conversational context and multimodal cues across encounters while letting older adults inspect, revise, and govern the memories through which family members participate.

\subsection{Relational Human--AI Interaction}

\citet{bickmore2005relationship} defined relational agents as computational artifacts designed to establish and maintain long-term social-emotional relationships with users. During repeated interactions at home, older adults have described these agents as friends \cite{bickmore2005acceptance_chi,bickmore2005acceptance}. \citet{vardoulakis2012companions} later tested a remotely operated in-home relational agent intended to support everyday conversation. Older adults frequently told stories about their lives during the week-long study, making remembering one of the most sustained topics even though it was not the agent's purpose \cite{vardoulakis2012companions}. Subsequent work identifies conversational breadth, appropriate responses, and continuity as important expectations \cite{vardoulakis2012companions,sidner2018companionable}, with sustained engagement supported by variation, self-disclosure, reflection, and relationship maintenance \cite{bickmore2010maintaining,skjuve2022relationships}. Nevertheless, older adults move between social and tool-oriented descriptions \cite{pradhan2019}, and conversational styles evoke close-relationship metaphors unevenly \cite{chin2024aunt}. Relationality can emerge through an agent's conduct and context without requiring users to believe it is human \cite{bickmore2005relationship,pradhan2019,chin2024aunt,satake2026refinement}.

Persistent memory can enact continuity, but how an agent recalls the past shapes that experience \cite{richards2014forgetmenot,cox2023previous,jo2024carecall,jiang2026recallbot}. Memoro inferred information needs from live conversational context and retrieved concise suggestions from recorded conversations, demonstrating context-sensitive retrieval \cite{zulfikar2024memoro}. Correct recall can strengthen believability and enjoyment, while errors produce frustration and weaken credibility \cite{richards2014forgetmenot}. References to earlier sessions can increase engagement, yet verbatim repetition can heighten privacy concerns \cite{cox2023previous}. CareCall similarly found that long-term memory supported familiarity and disclosure while raising topic-sensitive concerns around chronic conditions and privacy \cite{jo2024carecall}. RECALLbot, combining persistent memory, reciprocal disclosure, and user controls, increased perceived social identity, disclosure, and trust in a general-population sample \cite{jiang2026recallbot}. People with early-stage dementia valued MindTalker but wanted more consistent, personally meaningful interaction and emphasized that human relationships remained irreplaceable \cite{xygkou2024mindtalker}. Together, these studies favor accurate, relevant, selectively surfaced persistent memory under user control.

These challenges extend beyond the user-system dyad as voice agents enter multi-person homes and existing care relationships with distinct boundaries of ownership, access, and privacy \cite{luria2020boundaries,singh2026caregivers,karimi2025collaborative}. Agreeable responses can strengthen trust and empathy in low-stakes interactions \cite{mathur2026agreeableness}, while poorly timed or constrained responses can disrupt emotional pacing and user agency \cite{shi2026tensions}. This work connects the participatory study's (\S\ref{sec:pd}) Design Strategies concerning transparency, personalized follow-ups, continuity, and control to evaluation of remembered context, edit-and-delete controls, perceived empathy, and relational experience (\S\ref{sec:evaluation}).

Relational-agent research offers a way to sustain use beyond individual reminiscence encounters \cite{bickmore2010maintaining,skjuve2022relationships}. Memory companionship places remembering within an ongoing relationship, letting memories emerge through everyday conversation without making reminiscing the purpose of every interaction. Everyday remembering similarly begins in ordinary experiences, thoughts, and conversations, after which people decide whether to pursue, elaborate, or share what arose \cite{cosley2012everyday,blok2021facebook}. MeBo combines cues and conversational support from reminiscence research with continuity across encounters, while letting older adults inspect and correct retained memories and involve family members (\S\ref{sec:features}, \S\ref{sec:implementation}). This framing emerged through our participatory study, as older adults moved \emph{Memory Box} from a voice-based reminiscence intervention toward a companion (\S\ref{sec:pd}).

\section{How MeBo Came to Be}

Designing MeBo required connecting older adults' priorities with their experience of a working system. Participatory activities can elicit practices, values, and boundaries \cite{lindsay2012engaging,pradhan2020workshops,sakaguchi2021codesign}, while concrete artifacts support discussion of memory and emerging AI \cite{McNaney,maddali_investigating_2022}. Implementing these insights lets older adults assess design choices through interaction \cite{stegner2023situated,desai2023experience,gasteiger2022participatory,zhu2026reminiscope}. Evaluation then examined whether those choices reflected participants' priorities and revealed requirements that discussion alone had not surfaced.

MeBo follows this trajectory in three stages (Figure \ref{fig:teaser}). \emph{Finding MeBo} (\S\ref{sec:pd}) uses storyboards and participatory activities to develop four Design Strategies. \emph{Designing MeBo} (\S\ref{sec:designing}) translates these strategies into MeBo's features (Figure \ref{fig:features}). \emph{Evaluating MeBo} (\S\ref{sec:evaluation}) examines the functioning system through a temporally linked scenario sequence, testing usability, relational experience, and immediate emotional and social effects. This progression shows which participatory insights were carried into use, changed during implementation, or emerged only through interaction.


\section{Finding MeBo: A Participatory Study}
\label{sec:pd}

This section traces how \emph{Memory Box} became MeBo. Prior work guided the participatory study discussion: the memories and cues a system might support, how it might converse, what modalities and features it might include, and how to explore these with older adults. We analyzed how participants valued, combined, and bounded the possibilities these probes presented, yielding four Design Strategies. We call the study concept \emph{Memory Box} and introduce MeBo in DS1, where participants' responses shift from a memory-support tool toward a relational memory companion.

In June 2025, we searched the ACM Digital Library on older adults, cognitive impairment, memory support, and reminiscence, and voice- and AI-based conversational systems, retaining 38 papers offering relevant design or methodological insights. We sought enough prior work, rather than a systematic review, to scope the design space around \emph{Memory Box} and inform the participatory study.
One researcher deductively coded these with a codebook covering memory triggers and themes; system persona, modality, features, and contexts of use; study and evaluation methods; and reported successes, limitations, and interaction breakdowns \cite{Fife_Gossner_2024}. The team reviewed coding weekly over one month, resolving differences through negotiated agreement \cite{o2020intercoder}.
We developed five Design Considerations to represent probes, and four Participatory Design Considerations covering sessions' structure. Table~\ref{tab:design_considerations} shows the process toward four storyboards (S1--S4) and three exercises (EX1--EX3). \textbf{\textit{The search query, corpus, codebook, and coded matrix appear in the Supplementary Materials.}}

\begin{table*}[t]
\centering
\small
\caption{Design Considerations and Participatory Design Considerations derived from prior work, and how they informed the storyboard scenarios (S1--S4) and participatory exercises (EX1--EX3).}
\Description{A table listing five design considerations and four participatory design methods for a relational LLM-based voice reminiscence system, with descriptions, literature grounding, and the storyboard scenarios or participatory activities each informed.}
\label{tab:design_considerations}

\begin{tabularx}{\textwidth}{
p{0.8cm}
>{\RaggedRight\arraybackslash}p{2.6cm}
>{\RaggedRight\arraybackslash}X
p{1.4cm}
p{1.4cm}
}
\toprule
\textbf{ID} & \textbf{Name} & \textbf{Description} & \textbf{Grounding} & \textbf{Informs} \\
\midrule

\rowcolor{sectiongray}
\multicolumn{5}{@{}l}{\textbf{Design Considerations}} \\
\cmidrule(lr){1-5}

\textsc{DC1} & Intelligent Facilitation &
Structure conversation around appropriate themes and use open-ended questions, gentle prompts, and conversational scaffolds to deepen reminiscence without over-directing it.
& {\footnotesize \cite{yen_systematic_2018, hsieh_effect_2003, sun_chorus_2025}}
& S1, S4 \\

\addlinespace[4pt]

\textsc{DC2} & Longitudinal Context &
Maintain context across interactions by referencing prior conversations, life stories, and preferences to support continuity rather than isolated sessions.
& {\footnotesize \cite{duan_demo_2024, sun_chorus_2025}}
& S2, S4 \\

\addlinespace[4pt]

\textsc{DC3} & Multimodal Integration &
Integrate photographs, music, and other media as cues that are introduced, referenced, and followed up through conversation.
& {\footnotesize \cite{xu_memory_2024, herrera_empathetic_2024, wang_promoting_2024}}
& S2, S3, S4 \\

\addlinespace[4pt]

\textsc{DC4} & Emotional Complexity &
Account for the emotional sensitivity of reminiscence through appropriate responses, interpretive restraint, and user control over whether and how an interaction continues.
& {\footnotesize \cite{hsu_bittersweet_2025, lima_role_2023}}
& S1--S4 \\

\addlinespace[4pt]

\textsc{DC5} & Sensory Adaptation &
Accommodate sensory and interaction differences through accessible voice interaction, flexible pacing, simple presentation, and multimodal alternatives.
& {\footnotesize \cite{huang_designing_2025, cofre_voice_2020}}
& S1--S4 \\

\midrule

\rowcolor{sectiongray}
\multicolumn{5}{@{}l}{\textbf{Participatory Design Considerations}} \\
\cmidrule(lr){1-5}

\textsc{PDC1} & Evocative Grounding &
Begin with a reminiscence prompt and concrete storyboard scenarios so participants can respond to an instantiated interaction concept before discussing individual design features.
& {\footnotesize \cite{McNaney, maddali_investigating_2022}}
& Warm-up; S1--S4 \\

\addlinespace[4pt]

\textsc{PDC2} & Trigger Prioritization &
Use card sorting to identify meaningful memory themes and connect them with visual, auditory, and other reminiscence triggers.
& {\footnotesize \cite{Zhang, wang_promoting_2024}}
& EX1 \\

\addlinespace[4pt]

\textsc{PDC3} & Persona Curation &
Use adjective and archetype cards to surface preferred VUI traits, relational roles, communication styles, and conversational boundaries.
& {\footnotesize \cite{lima_role_2023, maddali_investigating_2022}}
& EX2 \\

\addlinespace[4pt]

\textsc{PDC4} & Feature Ideation \& Prioritization &
Use structured wishlist prompts followed by prioritization to move from open-ended ideas toward concrete capabilities participants considered most valuable.
& {\footnotesize \cite{McNaney, maddali_investigating_2022}}
& EX3 \\

\bottomrule
\end{tabularx}
\end{table*}

\subsection{Participants}
11 older adults (6 women, 5 men; 63–78 years, M=68.90, SD=5.07) from the U.S. participated in 90-minute remote co-design sessions. Nine participants were White or Caucasian and two Hispanic or Latino. All were native English speakers. Prior experience with conversational agents (CAs)\footnote{We use conversational agent as an umbrella term for AI systems people interact with through natural language, typed or spoken, and for the underlying conversational AI that produces those turns \cite{khadka2026empathy}.} varied. We recruited participants via university mailing lists and gave them a \$30 gift card. See participant demographics and prior CA experiences in Appendix, Table~\ref{tab:participants1}.

\subsection{Materials and Apparatus}

\begin{figure*}[t]
    \centering
    \includegraphics[width=\textwidth]{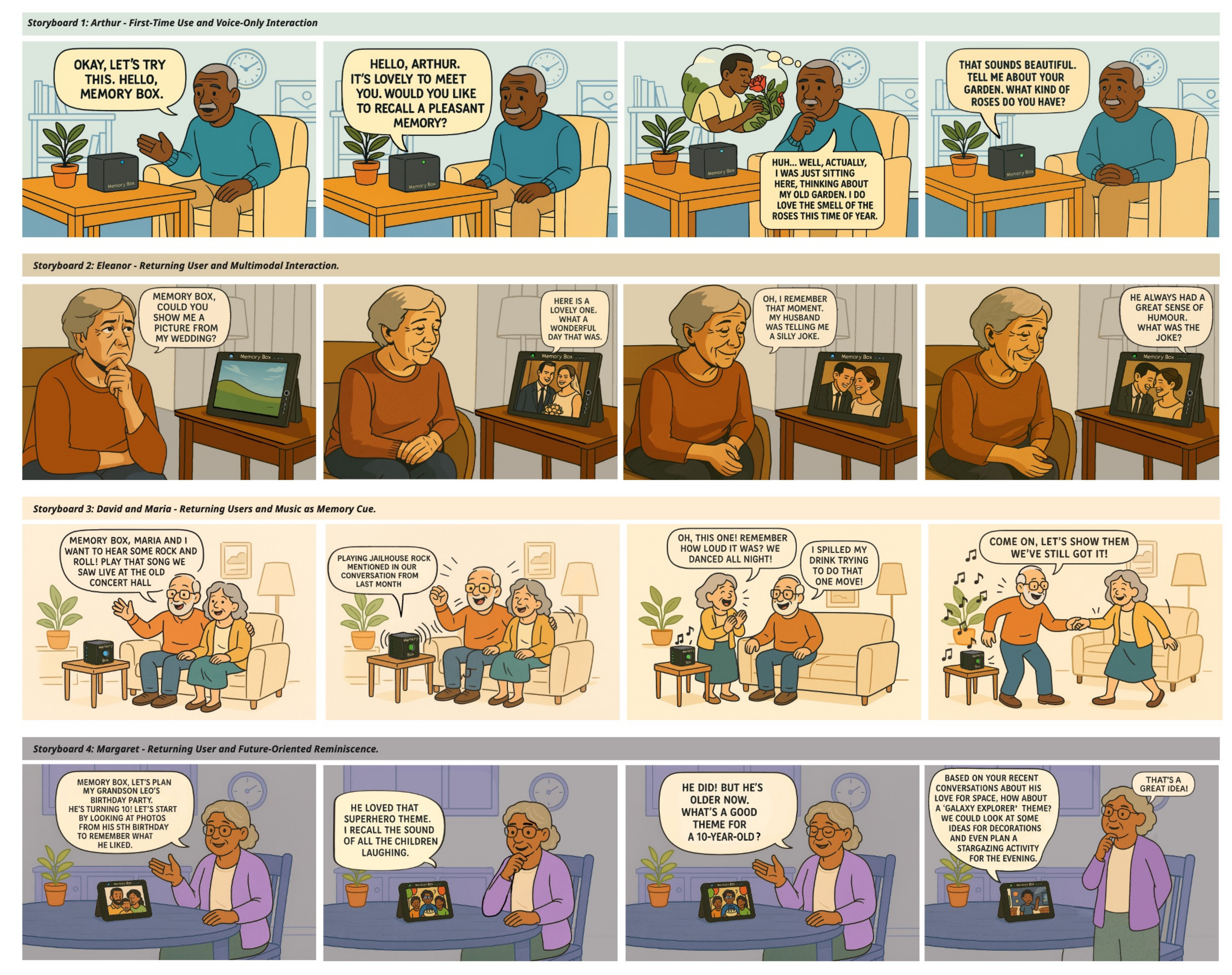}
    \caption{Four \emph{Memory Box} storyboards: (S1) first-time voice-only interaction; (S2) returning user with photographs; (S3) music-triggered reminiscence in reflective and shared contexts; and (S4) future-oriented reminiscence supporting intergenerational planning. Across scenarios, the system adapts to different user types, modalities, emotional contexts, and memory triggers.}
    \Description{Four rows of four comic panels show Memory Box use in domestic settings. In the first row, Arthur tries a voice-only device, receives a greeting and an invitation to recall a pleasant memory, talks about his old garden and roses, and receives a follow-up question about the roses. In the second, returning user Eleanor requests her wedding photograph; the system displays it, she recalls her husband's humorous remark, and the system asks what the joke was. In the third, David and Maria request a rock-and-roll song discussed previously. The system plays “Jailhouse Rock”; they recall dancing and spilling a drink, then stand and dance together. In the fourth, Margaret uses photographs and memories of her grandson's earlier birthday to plan his tenth birthday. The system suggests a space-themed party based on his recent interests, and Margaret agrees. Speech balloons and device screens carry the exchanges.}
    \label{fig:memorybox_storyboards}
\end{figure*}

\textbf{Storyboards.} We developed four low-fidelity storyboards as shared stimuli for discussing \emph{Memory Box}, varying user context, modality, memory cue, and temporal relationship with the system (Figure~\ref{fig:memorybox_storyboards}). They spanned first-time and returning use, individual and shared remembering, multimodal cues, and future-oriented and intergenerational interactions without prescribing a single interaction model. A muted palette, semi-abstract illustrations, and domestic settings focused attention on the interactions while letting participants project their own experiences.

\textbf{Collaborative Miro board.} We developed a Miro board to translate the design space into tangible materials for participatory exercises. In Exercise 1 (EX1), participants sorted and discussed cards representing memory themes and cues, including childhood, family, photographs, and songs. In EX2, participants selected and refined persona adjectives, relational archetypes, and conversational boundaries for the \emph{Memory Box}. In EX3, participants generated and prioritized capabilities through prompts such as ``I wish the \emph{Memory Box} could \ldots''. The board paired predefined materials with spaces for participants to add, rearrange, and modify ideas.

\subsection{Procedure}

\begin{figure*}[!htb]
  \centering
  \includegraphics[width=\textwidth]{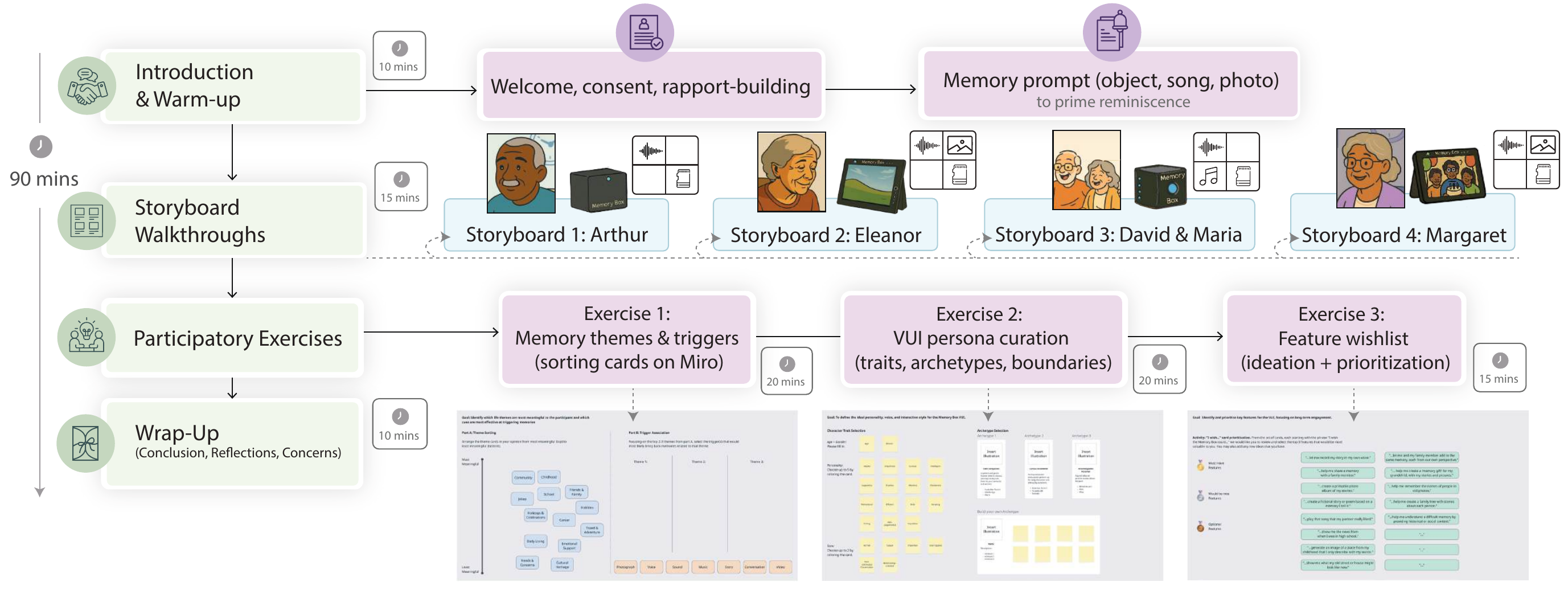}
  \caption{The four-phase, 90-minute participatory design session, including storyboard walkthroughs and three Miro-based exercises.}
  \Description{A flowchart illustrating the structure of a 90-minute co-design session in four phases. The first phase, Introduction and Warm-up, lasts 10 minutes and includes welcome, consent, rapport-building, and a memory prompt using an object, song, or photo to prime reminiscence. The second phase, Storyboard Walkthroughs, lasts 15 minutes and covers four storyboards: Arthur depicting voice-only, Eleanor depicting multimodal interaction, David and Maria depicting music as a memory cue, and Margaret depicting future-oriented reminiscence. The third phase, Participatory Exercises, consists of three exercises: Exercise 1, memory themes and triggers via card sorting on Miro, lasting 20 minutes; Exercise 2, VUI persona curation covering traits, archetypes, and boundaries, lasting 20 minutes; and Exercise 3, a feature wishlist through ideation and prioritization, lasting 15 minutes. The fourth phase, Wrap-Up, lasts 10 minutes, covers conclusions, reflections, and concerns. Screenshots of the Miro boards are shown beneath the exercise labels.}

  \label{fig:process-figure-memory-box}
\end{figure*}

We conducted individual, 90-minute co-design sessions via Microsoft Teams. Participants first signed an IRB-approved consent form and reported demographics, VUI experience, and use frequency. Figure \ref{fig:process-figure-memory-box} summarizes the four phases.

In \textbf{Phase 1: Introduction \& Warm-up}, we welcomed participants, established rapport, and introduced reminiscence and the workshop goals in nontechnical language. We used a memory prompt involving an object, song, or photograph to prime reminiscence. In \textbf{Phase 2: Storyboard Walkthroughs}, participants reviewed four low-fidelity storyboards illustrating interactions with the \emph{Memory Box} across user contexts, modalities, and memory triggers, including voice-only interaction, photographs, music, and socially situated memories. In \textbf{Phase 3: Participatory Exercises}, participants completed three structured activities on a collaborative Miro board: \emph{Exercise 1: Memory Themes \& Triggers}, sorting and discussing memory themes and evocative cues; \emph{Exercise 2: VUI Persona Curation}, selecting and refining personality traits, relational archetypes, and conversational boundaries; \emph{Exercise 3: Feature Wishlist}, generating and prioritizing features. Facilitators used prompts without directing participants toward particular outcomes, and we retained participants' board contributions as design artifacts. In \textbf{Phase 4: Wrap-Up}, participants reflected on their impressions, concerns, and boundaries for the \emph{Memory Box}, including privacy, control, emotional considerations, and its potential role in everyday and family life. We concluded with a debrief.

\subsection{Data Analysis}

Sessions were recorded in Microsoft Teams and transcribed. Following IRB protocol, data were securely stored, transcripts cleaned and anonymized, and recordings deleted after analysis. We employed thematic analysis \cite{Proudfoot_2023} using a hybrid inductive--deductive approach, letting patterns emerge from participants' discussions while remaining attentive to our research questions. Because sessions combined discussion with visual co-creation, we analyzed recordings and Miro artifacts together to preserve the context of participants' design decisions.

Two authors iteratively coded the dataset, tagging excerpts by participant and timestamp for traceability. Codes were developed reflexively and discussed with a senior author to resolve disagreements \cite{o2020intercoder}, then clustered into themes and reviewed against the dataset. From the broader themes, we report those directly informing the four Design Strategies. We describe prevalence as \textit{a few} ($\leq$20\%), \textit{some} (21--50\%), \textit{most} (51--80\%), and \textit{nearly all} ($>$80\%), indicating prevalence over importance. Study 1 identifiers are PA[X], where X is the participant number. \textbf{\textit{The codebook is available in the Supplementary Materials.}}

\subsection{Findings: Design Strategies}
\label{sec:strategies}

Besides feature preferences, participants described what relationship a reminiscence system should support, how memory cues should enter conversation, the social relationships within which remembering occurs, and what would keep the system trustworthy. We consolidated these into four Design Strategies (DS) that guided the subsequent design.

\subsubsection{DS1: Design for Relational Presence}

Participants encountered \emph{Memory Box} as a concept for reminiscence, but rarely evaluated it only as a tool for storing or retrieving memories. They considered whether it would listen, how it would respond, and whether interaction could feel like talking with someone rather than operating a device.
PA3 described the interaction as ``like having a little friend sitting there,'' valuing the open-endedness that lets one memory lead to another: ``if you talk about the garden, it can lead to another childhood memory or an old memory of the house.''

\textit{Nearly all} participants emphasized warmth, attentiveness, and responsiveness, describing their preferred persona. They favored familiar roles--- friend, sibling, or calm companion---over clinical or authoritative roles. \textit{Most} participants wanted the system to listen more than it spoke and let memories unfold without imposing a rigid sequence of questions. Conversational intelligence meant judging when to prompt, follow a memory, or remain quiet. As PA11 put it, the system should ``know to shut up and let them [characters in S3] dance.''

\begin{figure*}[t]
    \centering
    \includegraphics[width=\textwidth]{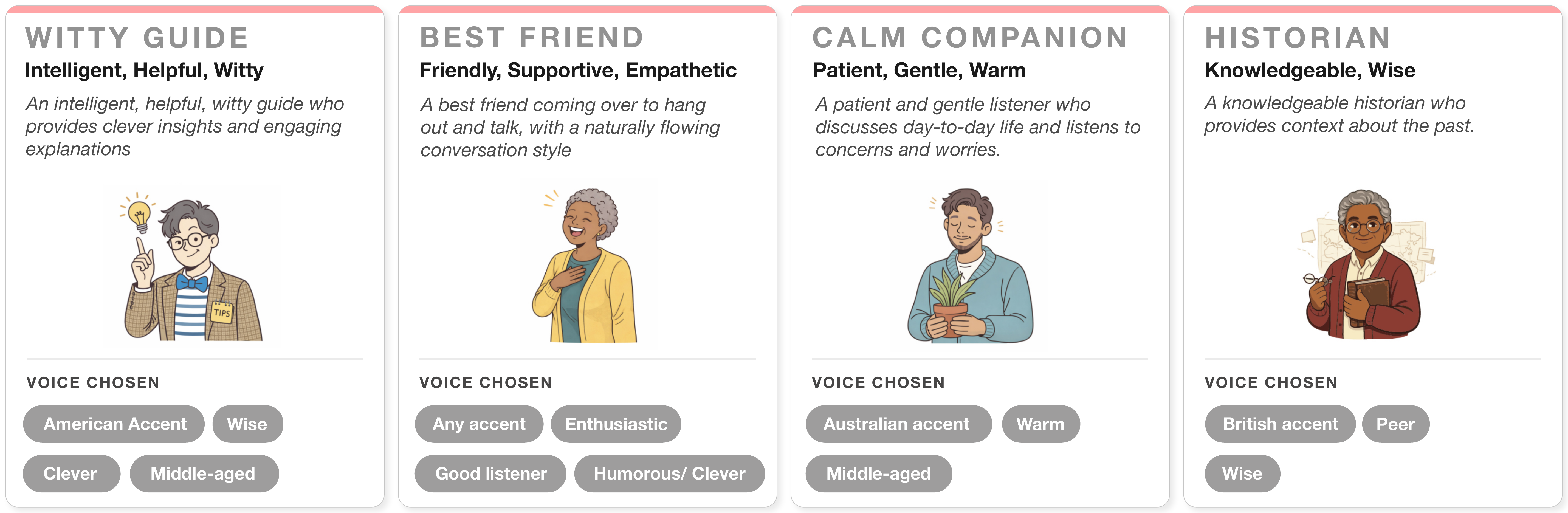} 
    \caption{Examples of personas developed during co-design sessions fused into four archetypes: a Witty Guide, a Historian, a Best Friend, and a Calm Companion.}
    \Description{A visual presenting four VUI persona archetypes developed during co-design sessions, each with key adjectives, an illustrated character, a short description, and voice characteristics. The first archetype, Witty Guide, is associated with the adjectives intelligent, helpful, and witty, illustrated as a middle-aged person in a blazer holding up a finger with a lightbulb, and described as an intelligent and witty guide who provides clever insights. The chosen voice is American-accented, middle-aged, wise, and clever. The second archetype, Best Friend, is associated with the adjectives friendly, supportive, and empathetic, illustrated as a cheerful person with a warm smile, and described as a best friend with a naturally flowing conversation style. The chosen voice is any accent, enthusiastic, a good listener, humorous, and empathetic. The third archetype, Calm Companion, is associated with the adjectives patient, gentle, and warm, illustrated as a calm person holding a potted plant, and described as a patient and gentle listener who discusses day-to-day life and concerns. The chosen voice is Australian-accented, middle-aged, and warm. The fourth archetype, Historian, is associated with the adjectives knowledgeable and wise, illustrated as an older person holding a book, and described as a knowledgeable historian who provides context about the past. The chosen voice is British-accented, peer-like, and wise.}
    \label{fig:personas-archetypes}
\end{figure*}

Persona-curation yielded four archetypes: Witty Guide, Best Friend, Calm Companion, and Historian (Figure~\ref{fig:personas-archetypes}).
These archetypes ranged from providing historical context to offering patient, naturally flowing conversation. Although their preferred traits and voices differed, each presented the system through a recognizable social role. 

Relational presence carried an expectation of emotional care. Participants described sharing memories as personal and vulnerable, making inappropriate enthusiasm, judgment, or excessive probing consequential. They wanted a conversational partner that was warm and non-judgmental without claiming to know a memory's meaning or feel it. They did not position the system as a replacement for human relationships. PA4 noted that ``it's almost sad talking to a box, and sharing your memories \ldots{}you don't have a person to share it with.'' Relational presence meant attentive, emotionally safe, bounded interaction.

This finding fundamentally changed how we framed the system. Rather than treating \emph{Memory Box} primarily as a container for memories---a box---DS1 positions it as a relational memory companion whose value depends on how it listens, responds, and remembers. We call this companion \textbf{MeBo}.

\subsubsection{DS2: Treat Memories and Cues as Conversational Anchors}

\begin{figure*}[h]
    \centering
    \includegraphics[width=\textwidth]{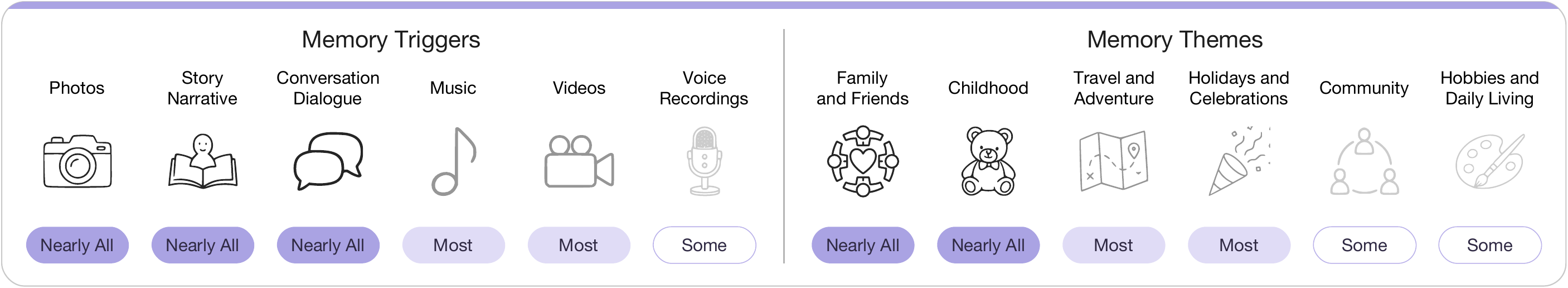} 
    \caption{Photographs, Story Narrative, and Conversation Dialogue emerged as the most universal triggers supporting memory recall. Memories tied to relationships and lived experiences were valued more than abstract emotional categories or functional needs.}
    \Description{A visual figure divided into two sections presenting participant preference rankings for memory triggers and memory themes. The left section, Memory Triggers, shows six trigger types ordered by preference: Photos, Story Narrative, and Conversation Dialogue are each rated Nearly All; Music and Videos are rated Most; and Voice Recordings is rated Some. The right section, Memory Themes, shows six theme types ordered by preference: Friends and Family and Childhood are each rated Nearly All; Travel and Adventure and Holidays and Celebrations are rated Most; and Community and Hobbies and Daily Living are each rated Some. Each item is accompanied by a small illustrative icon.}
    \label{fig:memory-triggers} 
\end{figure*}

Reminiscence emerged through memory cues combined with conversation. \textit{Nearly all} participants identified photographs, story narratives, and conversational dialogue as meaningful triggers. \textit{Most} valued music and videos, while \textit{some} identified voice recordings. Figure~\ref{fig:memory-triggers} summarizes these preferences alongside the most meaningful memory themes.

Participants did not separate an artifact from the conversation around it. A photograph mattered because it surfaced a person, place, or event that storytelling could then develop. As PA7 explained, ``old pictures always help with reminiscence \ldots{}if you look at an old picture, it lets you tell a story or think about that time.'' Music worked similarly, returning someone to a particular place and moment, such as feeling ``back in Portugal'' (PA11).
No single kind of cue was sufficient. Different cues could contribute to different memory elements and connect through conversation. PA6 summarized as ``music, photograph, video; conversations linked to those memories.''

DS2, therefore, treats memories and artifacts as conversational anchors: points from which MeBo can invite a story, follow its connections, and allow one memory to lead to another. The emphasis is not on providing the largest collection of media, but on supporting the conversation through which cues become meaningful.

\subsubsection{DS3: Support Socially Situated Memory}

\textit{Nearly all} participants prioritized memories involving friends, family, and childhood, while travel, celebrations, community, and hobbies were also valued (Figure~\ref{fig:memory-triggers}). Participants rarely discussed these memories as entirely private. They described remembering together through storytelling, comparing perspectives, and adding details. As PA11 explained, ``we’re doing this with my mom, but it’s the family doing it and not capturing the stories.'' Different versions of an event were not necessarily treated as mistakes; participants recognized that ``history can be retold many ways'' (PA3).

Participants also considered how recorded memories might reach absent family members. \textit{Most} valued preserving stories in the speaker’s own voice and turning them into tangible or shareable artifacts. More than archival storage, these possibilities were tied to legacy and intergenerational connection:

\begin{quote}
\textit{``It would be wonderful that when the person passes, those stored memories and conversations don’t go away. Someone can say, I now have these reminiscences. There are stories from my grandfather, and I can listen to his own voice talking and pass that reminiscence down the road.''} (PA5)
\end{quote}

Participants consequently imagined family members contributing photographs, adding perspectives, or receiving stories and recordings. Printable albums and other tangible outputs were valued for letting memories move beyond the conversation. Simultaneously, concerns about ``talking to a box'' cautioned against designing reminiscence as a substitute for human contact. DS3 therefore situates MeBo within existing relationships, supporting co-created memories, family participation, and legacy while preserving the user’s control over what is shared.

\subsubsection{DS4: Build Continuity with Transparency and Control}

Participants evaluated \emph{Memory Box} with repeated use in mind, expecting later conversations to draw on accumulated history instead of starting anew. PA2 made this assumption explicit: ``OK, so you’re assuming I’ve been talking to this thing for a long time, and it has some history with me, right? I wouldn’t have that whole conversation.'' Recalling earlier stories, preferences, and conversational context was how participants distinguished an ongoing companion from a tool.

Continuity, however,  made the stored information more sensitive. Participants described memories as personal and vulnerable, and raised concerns about where they would be stored and how they might be used:

\begin{quote}
\textit{``As long as it's not connected to an online AI. I know if you're connected to an AI device, that device will eventually collect data \ldots{}So when you're dealing with personal memories, I would really think about data.''} (PA3)
\end{quote}

Participants wanted control over when the system engaged, what it sensed, and who could access the resulting information. They objected to always-on or opaque interaction and preferred to initiate conversations themselves: ``I wouldn’t want the device to engage me, I would want to engage the device'' (PA4). They also wanted clear indications of AI-generated content and visible ways to restrict access to personal memories.

\begin{figure*}[!htb]
\centering
\includegraphics[width=\textwidth]{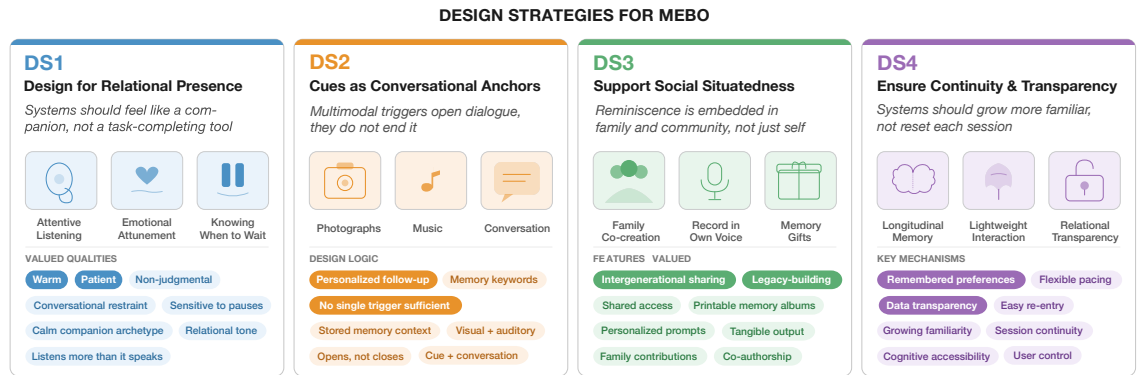}
\caption{The four Design Strategies derived from the participatory study. The figure summarizes the qualities, design logic, desired features, and interaction mechanisms through which participants framed MeBo as a relational, conversational, socially situated, and longitudinal memory companion.}
\Description{A four-column summary of the Design Strategies for MeBo. DS1, Design for Relational Presence, emphasizes warmth, patience, non-judgment, attentive listening, emotional attunement, conversational restraint, and knowing when to wait. DS2, Treat Memories and Cues as Conversational Anchors, shows photographs, music, and conversation as connected cues that open dialogue through personalized follow-up and stored memory context. DS3, Support Socially Situated Memory, emphasizes family participation, co-creation, recordings in the user’s own voice, intergenerational sharing, legacy-building, shared access, and tangible memory artifacts. DS4, Build Continuity with Transparency and Control, emphasizes longitudinal memory, remembered preferences, flexible pacing, easy re-entry, data transparency, user control, and continuity across sessions.}
\label{fig:ds}
\end{figure*}

DS4, therefore, couples longitudinal continuity with transparency and control. MeBo must remember to become familiar over time; users must be able to govern that memory for familiarity to remain acceptable. Continuity is not simply the technical retention of conversational history. It is an arrangement in which remembered information remains visible, bounded, and under the user’s direction. Figure \ref{fig:ds} summarizes design strategies.

\section{Designing MeBo: A Voice-Based Memory Companion}
\label{sec:designing}
\begin{figure*}[!htb]
\centering
\includegraphics[width=\textwidth]{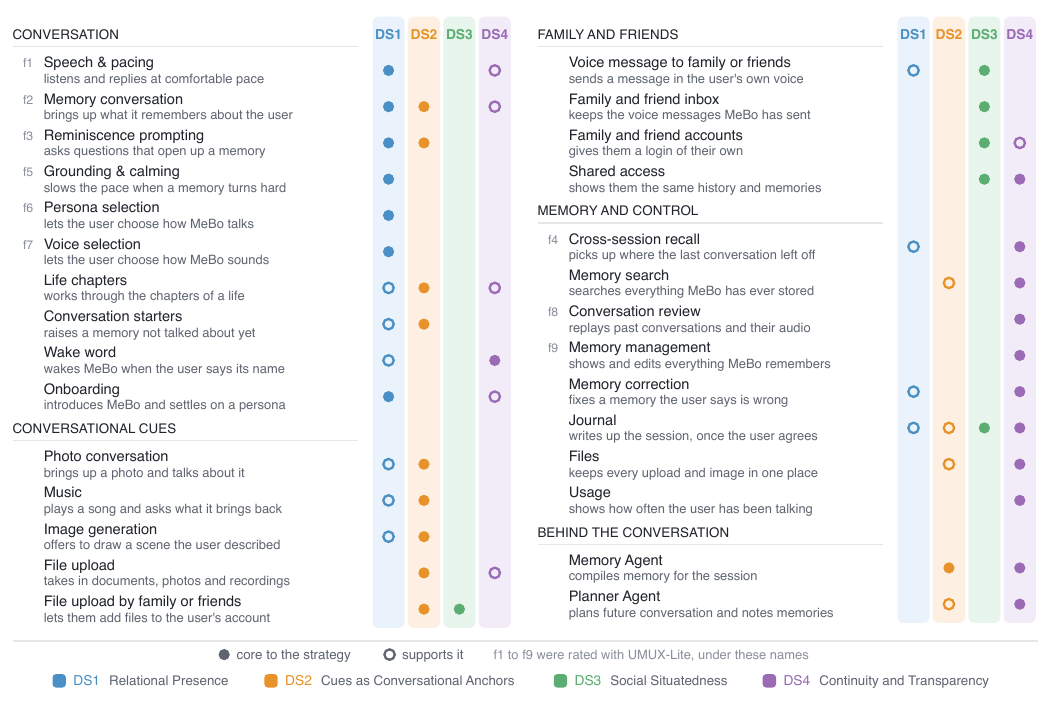}
\caption{MeBo's feature set, derived from the four Design Strategies. A filled circle marks a feature core to a strategy; a hollow circle marks one that supports it.}
\Description{A two-column map of MeBo's features against the four design strategies. Rows are grouped into five bands. Conversation holds speech and pacing, memory conversation, reminiscence prompting, grounding and calming, persona selection, voice selection, life chapters, conversation starters, wake word, and onboarding. Conversational cues include photo conversation, music, image generation, and file uploads by family or friends. Family and friends can send voice messages to family or friends, use a family and friend inbox, have family and friend accounts, and share access. Memory and control include cross-session recall, memory search, conversation review, memory management, memory correction, journal, files, and usage. Behind the conversation are the Memory Agent and the Planner Agent. Each row carries four marks, one per strategy: a filled circle indicates the feature is core to that strategy, and a hollow circle indicates it supports it.}
\label{fig:features}
\end{figure*}

The four Design Strategies from \emph{Finding MeBo} (\S\ref{sec:pd}) bridge participants' accounts and the design of MeBo. Guided by these strategies, the design and development teams iteratively finalized MeBo's features in meetings. We included a feature only if it was core to at least one strategy or clearly supported one. Specific findings also shaped MeBo. Participants' memory themes became the ten life chapters, and memory triggers---photos, music, and places---became conversational cues. The relational qualities they valued informed MeBo's persona and voice selection, its speech and pacing, and how it prompts reminiscence. Their concerns about longitudinal use informed cross-session recall, memory search, conversation review, memory management, and correction. Figure \ref{fig:features} maps MeBo's features to strategies. We describe interaction design by strategy (\S\ref{sec:features}), then the supporting architecture (\S\ref{sec:implementation}).

\subsection{Interaction Design}
\label{sec:features}
MeBo is a voice-based memory companion for older adults. Our design strategies (\S\ref{sec:strategies}) call for an interaction organized around conversation about older adults' lives rather than a general-purpose assistant. MeBo, therefore, asks about their lives, listens, and uses what it learns to connect later conversations to earlier ones. MeBo's conversational behavior draws on reminiscence practice \cite{yen_systematic_2018, hsieh_effect_2003}: attending to sensory detail, staying with one topic, and building follow-up questions from the user's account. 

\subsubsection{Relational Conversation}
DS1 calls for the conversation itself to feel relational. MeBo adopts one of four personas from the participatory sessions (Figure \ref{fig:personas-archetypes}), selected independently of the voice. We curated two male and two female voices. Across personas, MeBo gives a brief reaction and asks one question at a time, using sensory details to guide users in remembering and narrating their memories. We designed MeBo to support remembering without dictating what a memory means or how significant it should be.

\subsubsection{Memories and Conversational Cues}
MeBo organizes what it learns into ten personalized life chapters. Following DS2, MeBo treats these memories as conversation openings. When a conversation needs a new direction, MeBo introduces a less-explored chapter and draws on related stored information. Photographs, music, places, and uploaded artifacts serve as cues. Rather than simply presenting a cue, MeBo connects it to associated people, places, and events, then asks sensory or experiential questions. Users and, with permission, friends and family members can contribute cues for later conversations (DS3).

\subsubsection{Family and Social Memory}
DS3 situates memories in relationships while preserving the user's control. Users can grant relatives or friends view or edit access to their conversations, journal, files, and memories, and revoke that access any time. People with edit access can contribute photographs and correct details, shaping what MeBo later recalls. Users can also record memories or messages in their voice for family and friends. 

\subsubsection{Continuity, Transparency, and Control}
DS4 makes continuity acceptable only when users can see and govern what MeBo remembers. Each session resumes the previous session's unfinished thread. MeBo surfaces open worries and due follow-ups before lower-priority similarity search results. Its dashboard exposes accumulated memory and labels each fact's source, separating what the user stated (\emph{You said this}) from what MeBo inferred (\emph{MeBo noticed}). Users can correct memories during voice conversations or edit and delete them in the dashboard. For added control, MeBo uses a custom browser-based wake-word detector trained by our team. These interaction decisions, guided by the design strategies, define how MeBo behaves in a conversation.

\subsection{System Architecture and Implementation}
\label{sec:implementation}
\begin{figure*}[!htb]
\centering
\includegraphics[width=.85\textwidth]{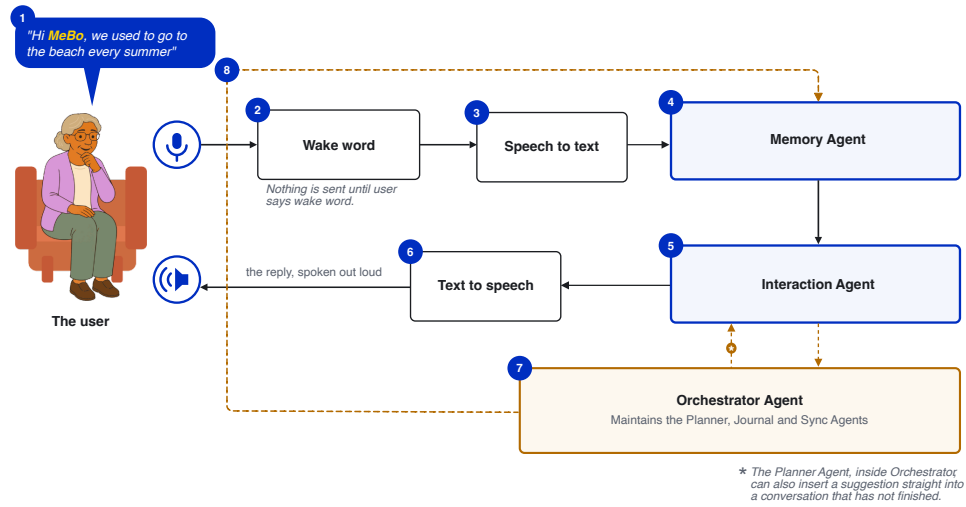}
\caption{MeBo's interaction loop, numbered in order. Older adults wait on steps 1--6 only. Steps 1--6 happen in real-time. Step 7 runs in the background, and step 8 is MeBo's preparation for future turns. Blue marks a language model call.}
\Description{A diagram of MeBo's processing loop drawn as a cycle. On the left, an older adult in an armchair speaks the sentence ``Hi MeBo, we used to go to the beach every summer", labelled step one. Her speech passes through a microphone to step two, wake word detection, annotated "nothing is sent until user says wake word", then step three, speech to text, then step four, the Memory Agent, and step five, the Interaction Agent, both drawn in blue to mark a language model call. The reply returns through step six, text-to-speech, to a speaker beside her, labeled "the reply, spoken out loud". A dashed amber line carries the finished turn down to step seven, the Orchestrator Agent, which maintains the Planner, Journal, and Sync Agents, and a dashed amber lane labeled step eight runs from there back into the Memory Agent for the next turn. A starred amber arrow back into the Interaction Agent notes that the Planner Agent can also put a suggestion straight into a conversation that has not finished.}
\label{fig:system-overview}
\end{figure*}

The architecture enables these interaction decisions in real time. MeBo's architecture has two core requirements. First, MeBo must retrieve enough context to respond during the live turn. Second, MeBo can extract memories and plan for future conversations in the background. The architecture, therefore, separates latency-sensitive from deferred processes (Figure \ref{fig:system-overview}).

\subsubsection{Live Conversation}

Imagine a user says, \emph{``Hi MeBo, we used to go to the beach every summer.''} Wake-word detection runs locally, detecting \emph{``MeBo''} before any audio is transmitted. The utterance passes through speech recognition and turn detection into the live loop, where barge-in lets the user interrupt a reply. Two agents---Memory Agent and Interaction Agent---run in this loop. The Memory Agent performs the only retrieval the user waits for. Each turn, it assembles an ordered memory packet: unresolved matters, the current conversation, relevant cues and memories, and response guidance (Figure \ref{fig:memory}). This ordering lets a week-old worry about a sore knee surface before older matches to the beach just mentioned. The Interaction Agent makes one language model call over the persona, conversation guide, memory packet, and any current nudge (a greeting, due follow-up, Planner suggestion, or topic change) to produce a reply. The Interaction Agent detects turn endings with a turn-completion classifier, and can invoke tools for memory, media, preferences, or journaling (Figure \ref{fig:interaction}).

\subsubsection{Background Memory Processing}
\begin{figure*}[!tb]
\centering
\includegraphics[width=\textwidth]{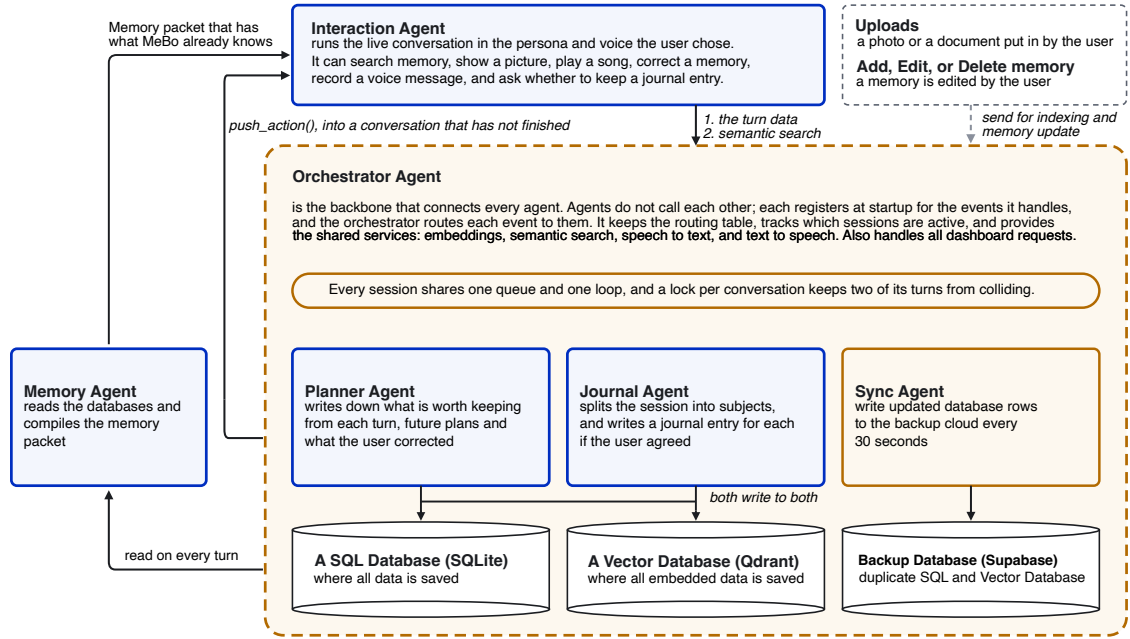}
\caption{How MeBo's agents work together. Interaction and Memory run in the live conversation; Planner, Journal, and Sync run inside the Orchestrator, after the person has been answered. Agents never call each other: each registers at startup for the events it handles, and the Orchestrator routes them.}
\Description{A diagram of MeBo's agent architecture. At the top, the Interaction Agent runs the live conversation in the persona and voice the user chose, and can search memory, show a picture, play a song, correct a memory, record a voice message, and ask whether to keep a journal entry. To its right, uploads and user edits are sent to memory for indexing. A dashed amber boundary encloses the Orchestrator Agent. Inside it are the Planner Agent, which records what is worth keeping from each turn; the Journal Agent, which splits the session into subjects and writes an entry for each if the user agrees; and the Sync Agent, which copies updated rows to the backup cloud. Below them sit a SQL database, a vector database, and the cloud backup. Outside the boundary on the left, the Memory Agent reads the databases on every turn and compiles the memory packet passed to the Interaction Agent, and an arrow from the Planner Agent runs back into a conversation that has not finished.}
\label{fig:agent-map}
\end{figure*}

The Orchestrator queues each completed turn for background processing, without delaying the next conversation (Figure \ref{fig:agent-map}). The Planner Agent extracts candidate facts, concerns, cues, plans, and verbatim user statements, comparing each with existing memory to distinguish new information from duplicates and possible corrections (Figure \ref{fig:planner}). Users can edit or erase anything it stores. With the user's consent, the Journal Agent organizes a conversation into topics and summarizes each in the user's own words, merging related material into existing entries so recurring memories accumulate. Later edits re-enter the pipeline and change what MeBo says aloud. The agents do not call one another. Instead, the Orchestrator routes between them, limiting any failure to a single turn's memory. Memory extraction combines retrieval-augmented generation (RAG) with a language model. Each agent writes to the two primary storage first; the Sync Agent then mirrors changes to the cloud in the background.

\begin{figure*}[t]
  \centering
  \includegraphics[width=\textwidth]{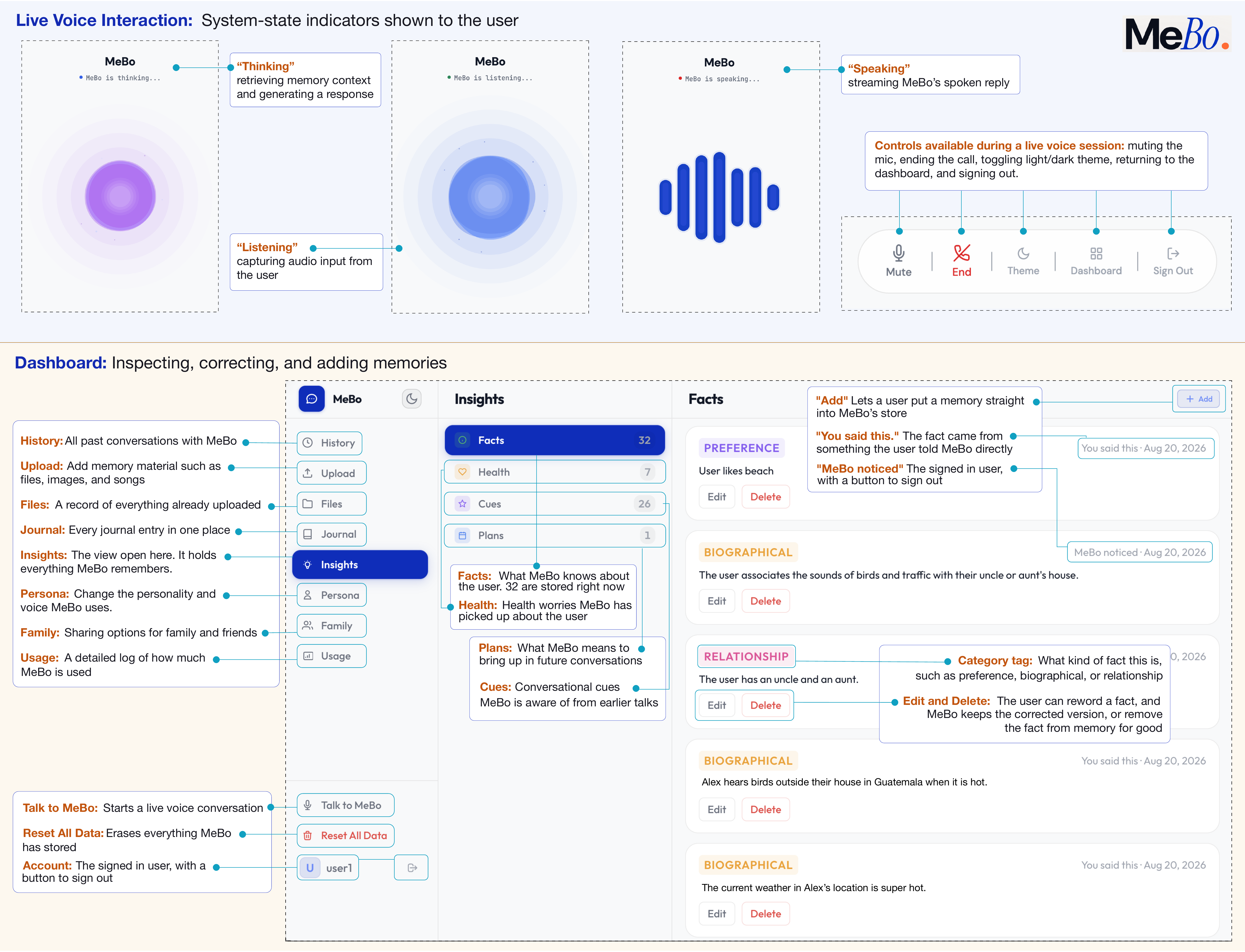}
  \caption{MeBo's interface: (Top Panel) Live Voice Interaction; (Bottom Panel) The Insights dashboard displaying the system-state indicators shown to the user while interacting with MeBo}
  \Description{A two-part screenshot of the MeBo interface. The Top Panel shows the Insights dashboard with a left navigation sidebar, a list of insight categories (Facts, Health, Cues, Plans), and a list of fact cards, each showing a category tag, the fact text, provenance (whether the user stated it or MeBo inferred it), and edit/delete controls. The Bottom Panel shows three states of the voice interface (listening, thinking, and speaking) each represented by an animated orb or waveform icon.}
  \label{fig:ui-diagram}
\end{figure*}

\subsubsection{Implementation}
MeBo runs in the browser with a FastAPI\footnote{\url{https://fastapi.tiangolo.com/}} backend. We use SQLite\footnote{\url{https://www.sqlite.org/}} and an embedded Qdrant\footnote{\url{https://qdrant.tech/documentation/}} vector index as the primary databases; a separate Supabase\footnote{\url{https://supabase.com/docs/}} instance serves as a cloud backup. 


MeBo supports different models and vendors. Every model uses a common interface, letting the language model, embedding model, speech-to-text, and text-to-speech providers be swapped without changing MeBo's agents. The evaluation configuration uses Gemini~3 Flash\footnote{\url{https://ai.google.dev/gemini-api/docs/models/gemini-3-flash-preview}} for conversation, memory extraction, and journaling; Qwen3-Embedding-4B\footnote{\url{https://qwenlm.github.io/blog/qwen3-embedding/}} for semantic retrieval; Llama~4 Scout\footnote{\url{https://ai.meta.com/llama/get-started/}} for uploaded-file description; Deepgram Nova-3\footnote{\url{https://developers.deepgram.com/docs/models-languages-overview}} for speech-to-text; and Cartesia\footnote{\url{https://docs.cartesia.ai/get-started/overview}} for text-to-speech. Together, these design and implementation decisions instantiate the four strategies as a functioning system for evaluation.

\section{Evaluating MeBo: A User Study}
\label{sec:evaluation}

\emph{Finding MeBo} produced four Design Strategies for relational memory companions: relational presence (DS1), cues as conversational anchors (DS2), social situatedness (DS3), and continuity and transparency (DS4). We implemented these strategies in MeBo and evaluated its usability and feature value, relational experience and acceptance, and immediate emotional and social effects.

Evaluating remembering and several defining qualities of MeBo posed a methodological challenge because they unfold over time. Participants in the participatory study situated remembering within everyday life, ongoing storytelling, and relationships, while DS4 called for accumulating memories, preferences, and conversational history. They associated a system revisiting remembered interactions with a companion, and one repeatedly resetting with a tool. Evaluation therefore required memories to accumulate, be revisited, and connect across interactions---conditions that cannot naturally develop within a time-limited session.

We made these longitudinal interactions imaginable within one session through a vignette-based scenario approach, drawing on vignette methodology \cite{finch1987vignette,barter1999use,aguinis2014best,atzmuller2010experimental} and its use in HCI \cite{van2025unhealthy,von2023attendant}, particularly with older adults \cite{mathur2026facts}. Vignettes allow participants to fill unspecified details from their own experiences, while continuous narratives accumulate context across events \cite{Hughes_1998,Hughes_Huby_2004}. Scenario-based approaches have similarly structured older adults' interactions with voice assistants and conversational agents for social engagement \cite{Joshi_Ulabhaje_Nataraj_Martin-Hammond_2025}. We extended this approach by linking eight scenarios across approximately one year, letting shared context accumulate as participants interacted with the functioning MeBo.

The scenarios followed Alex Mitchell, a fictional 65-year-old user with a gender-neutral name \cite{iacovides2022close}, from receiving MeBo through approximately one year of use. Participants adopted Alex's situation while drawing on their own memories. We held Alex's basic biography and timeline constant to establish a shared history that MeBo could revisit while leaving space for participants' experiences. A four-item continuity measure assessed whether the scenarios conveyed connected moments, a shared history with MeBo, and continuity extending across a year; ratings confirmed that the sequence conveyed the intended passage of time (\S\ref{sec:evaluation}). Accordingly, we asked:

\textbf{-RQ1 (Usability and Feature Value):} How do participants perceive the usability of MeBo, and which features do they find most useful?

\textbf{-RQ2 (Relational Experience and Acceptance):} How do participants experience MeBo as a conversational and relational partner, and what factors shape their acceptance of it?

\textbf{-RQ3 (Immediate Emotional and Social Experience):} How does interacting with MeBo affect participants' immediate emotional state and sense of social connection, and what do participants attribute these effects to?

\subsection{Participants}

\begin{figure}[htbp]
    \centering
    \includegraphics[width=0.49\textwidth]{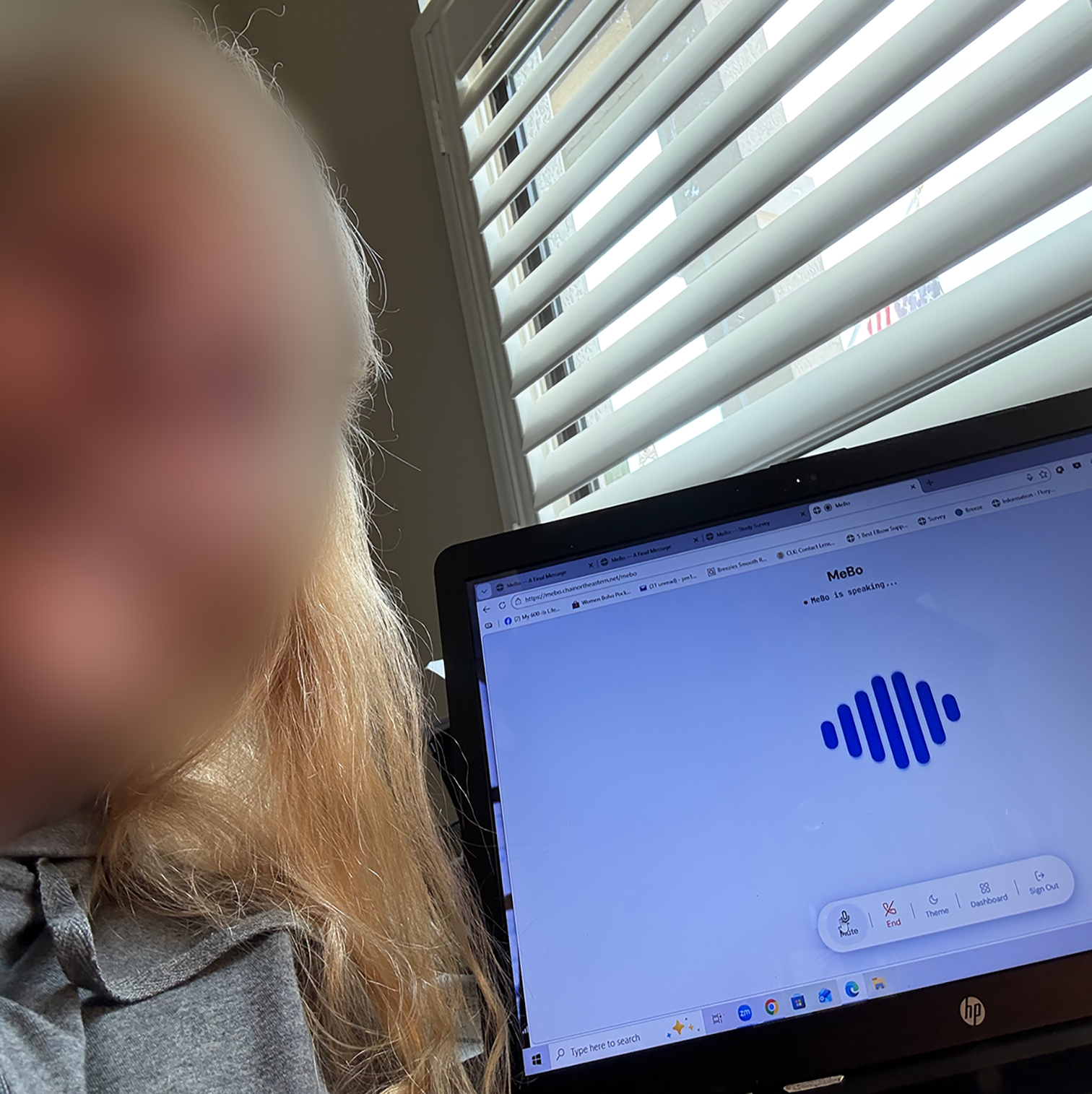} 
    \hfill
    \includegraphics[width=0.49\textwidth]{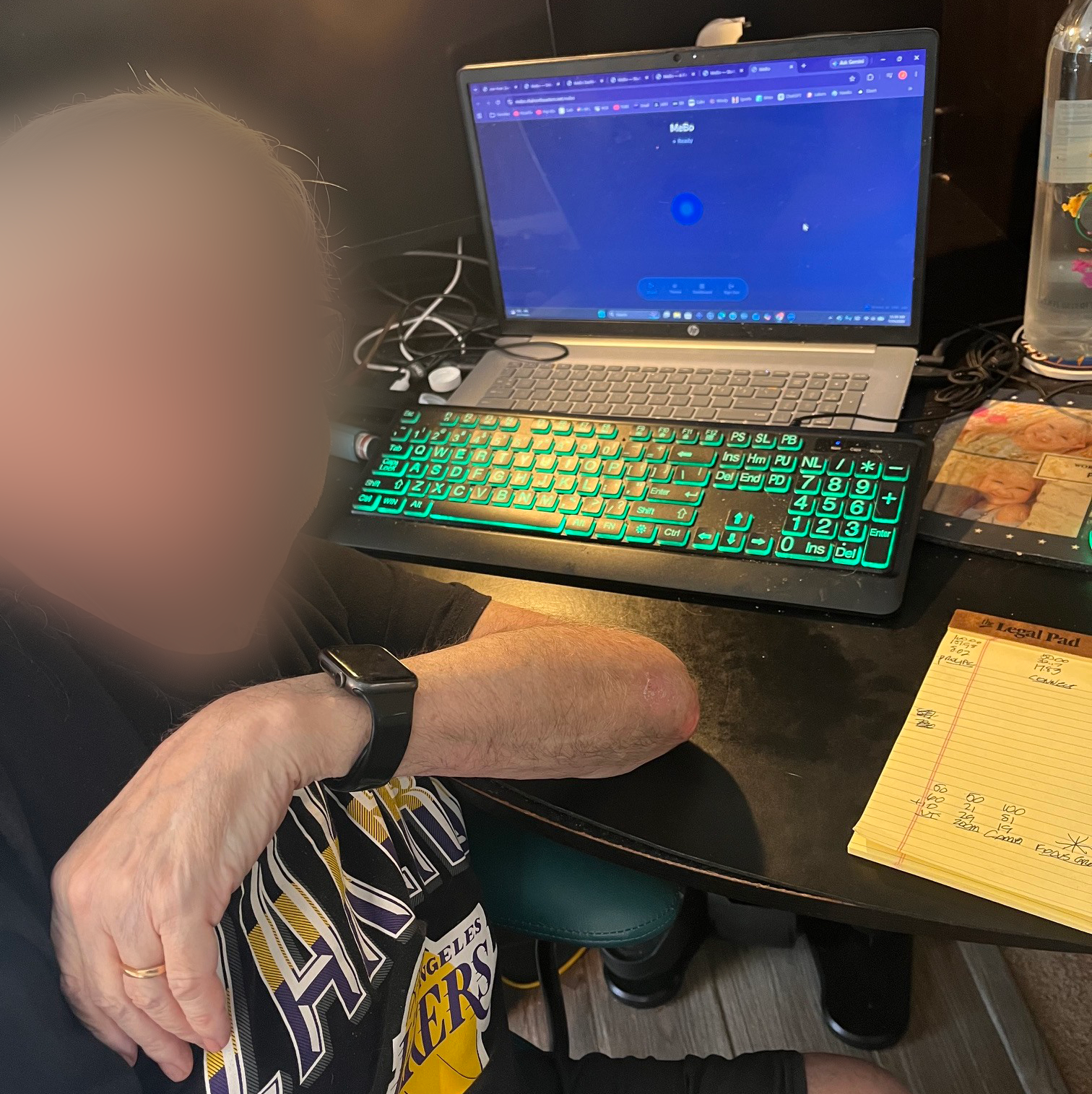} 
    
    \Description{Two side-by-side photographs show participants beside laptops running MeBo; both participants' faces are blurred. On the left, a participant sits beside a window and a laptop showing a blue speaking waveform on the light interface. On the right, a participant sits at a desk with a laptop and an external keyboard; the laptop shows a blue orb on the dark interface.}
    \caption{Two participants interacting with MeBo in light mode (left) and dark mode (right).}
    \label{fig:participants_with_MeBo}
\end{figure}

20 older adults (11 women, 9 men; 66-78 years, $M$=70.50, $SD$=3.68) from the U.S. participated in 90-minute remote sessions (Fig. \ref{fig:participants_with_MeBo}). The sample included 17 White or Caucasian participants, two Black or African American participants, and one Native American participant, all native English speakers. 
Prior CA experience varied; all but one were familiar. We recruited through Prolific\footnote{https://www.prolific.com/} to broaden participation beyond the university networks that had produced a homogeneous, highly educated sample in the participatory study. Participants received \$30 after study completion. See details on demographics and CA experiences in Appendix Table~\ref{tab:participants_S2}.


\subsection{Materials and Apparatus}

\textbf{MeBo.} Participants used the functional MeBo system described in Section~\ref{sec:pd} through its browser-based interface. Interaction occurred primarily through a voice call, with scenarios involving photographs, music, voice messages, journaling, cross-session recall, and MeBo's dashboard. 

\textbf{Pre-populated longitudinal history.} For the simulated year, we pre-populated each participant's MeBo account with plausible fictional autobiographical disclosures from Alex: family information, a garden photograph, and a wedding song. This lets later scenarios exercise MeBo's actual memory-retrieval and resurfacing mechanisms. Scenarios relying on this shared history constrained some elements of Alex's biography, while others remained open for participants' own memories.

\begin{figure*}
    \centering
    \includegraphics[width=\textwidth]{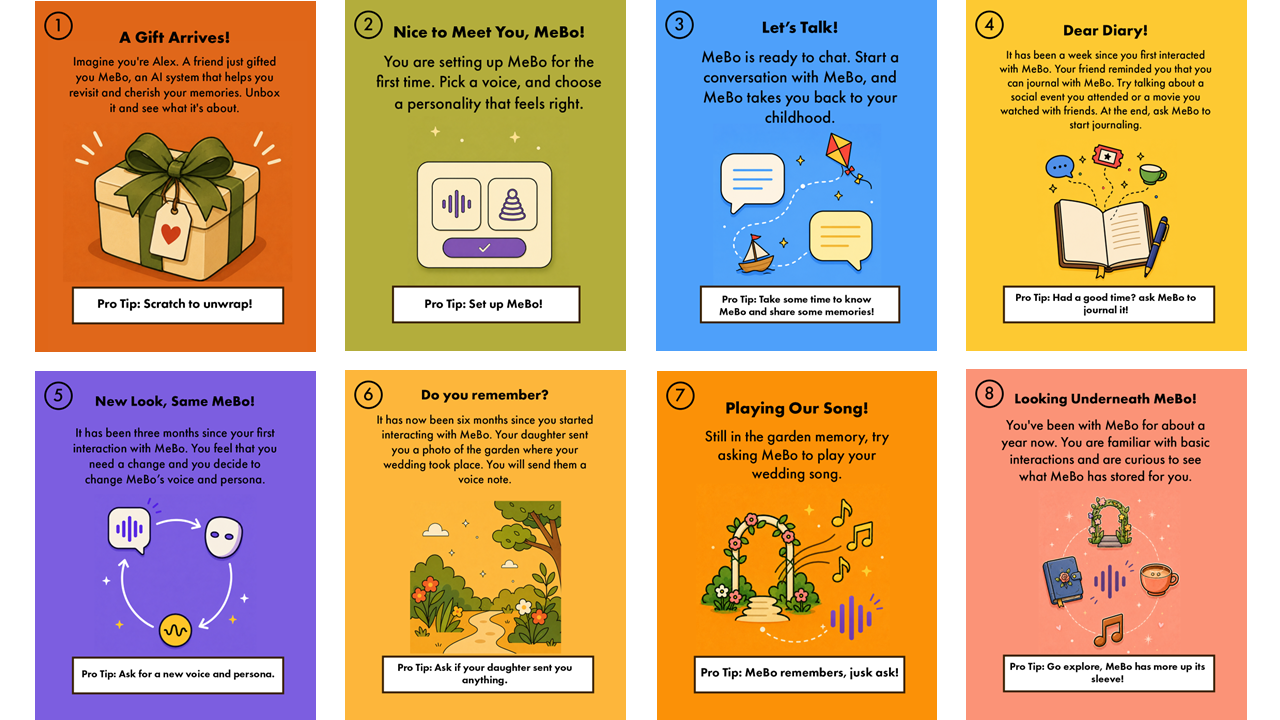}
    \caption{The eight scenario cards, representing a task in Alex's simulated timeline with MeBo, from first contact to one year. Cards were presented one at a time in fixed order, with an optional Pro Tip for participants who needed guidance. }
    \Description{Eight numbered scenario cards in two rows of four, each in a different color, showing a title, scenario text, an illustration, and a Pro Tip. The cards move from receiving MeBo as a gift and setting it up, through conversation, journaling, and changing its voice, to responding to a family photo, playing a wedding song, and exploring stored memories after a year.}
    \label{fig:cards}
\end{figure*}

\textbf{Scenario cards.} Eight scenario cards presented the simulated year in fixed order, each situating participants in Alex's relationship with MeBo and requesting an interaction (Figure~\ref{fig:cards}). The sequence extended from first use to approximately one year, letting later scenarios depend on context accumulated earlier. Together, the cards operationalized the four DS through voice and persona selection (DS1); conversational use of photographs and music (DS2); journaling, family sharing, and legacy (DS3); and accumulated memory, transparency, and control through the dashboard (DS4). The team created visually distinct cards using ChatGPT Images \footnote{ChatGPT Images 2.0, accessed via \url{https://chatgpt.com}}, iteratively refined their content and sequencing with input from a member experienced in visual design, and informally piloted them with university peers and older adults in our research networks.

\subsection{Measures}
\subsubsection{Usability and Feature-Level Evaluation}
\label{sec:measures-usability}
To answer RQ1, we assessed overall usability with the 10-item System Usability Scale (SUS; five-point items; 0--100 score) \cite{brooke1996sus}. Since SUS yields one system-level score, we also assessed the nine directly exercised features in Figure \ref{fig:sus}B by adapting both UMUX-Lite items---capability and ease of use---to each feature (seven-point items) \cite{lewis2013umuxlite}.


\subsubsection{Relational Experience and Acceptance}
\label{sec:measures-relational}
To answer RQ2, we used three Almere subscales developed for older adults' acceptance of social agents: intention to use (three items), perceived enjoyment (five), and perceived sociability (four), each rated from 1 (strongly disagree) to 5 (strongly agree) \cite{heerink2010almere}. We measured perceived empathy with the 10-item Perceived Empathy of Technology Scale (PETS): emotional responsiveness (six items) and understanding and trust (four), rated on 0--100 sliders \cite{schmidmaier2024pets}. We averaged items within each subscale; higher scores indicate more positive perceptions.

\subsubsection{Affect and Social Connection}
\label{sec:measures-affect}
For RQ3, participants completed all measures immediately before and after MeBo interaction. The 20-item Positive and Negative Affect Schedule (PANAS) measured positive and negative affect as two 10-item sums (1--5 per item; 10--50 per subscale) \cite{watson1988panas}. The three-item UCLA Loneliness Scale measured perceived loneliness (1--3 per item; summed to 3--9) \cite{hughes2004loneliness}. We averaged two study-specific seven-point items measuring momentary social connection (1--7) to distinguish immediate connection from loneliness.

We also measured whether participants perceived continuity across eight scenario cards simulating a year of use. Participants rated four items (1--7, strongly disagree to strongly agree): the experience helped them imagine a year with MeBo, the scenarios felt like connected moments, they and MeBo had developed a shared interaction history, and MeBo's stored memories created continuity across the scenarios. We averaged items into one continuity score.

\subsection{Procedure}

\begin{figure}[t]
    \centering
    \includegraphics[width=\linewidth]{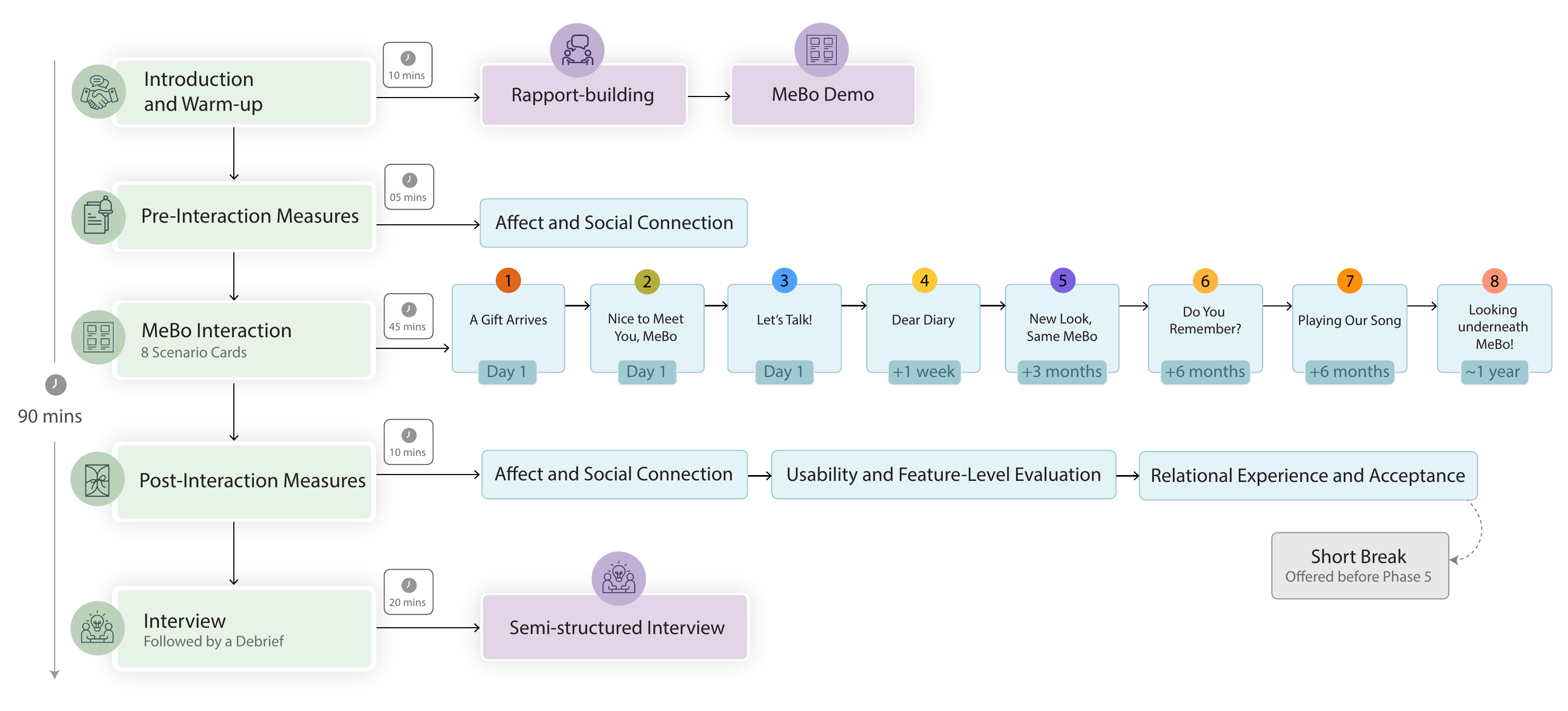}
    \caption{Study 2 Procedure — Evaluating MeBo. Zoom sessions (consented, recorded \& transcribed) \(\cdot\) N = 20 older adults recruited via Prolific.}
    \Description{A flowchart lays out five phases of a 90-minute evaluation. Introduction and Warm-up takes 10 minutes for rapport-building and a MeBo demonstration. Pre-Interaction Measures takes 5 minutes for affect and social connection. MeBo Interaction takes 45 minutes and expands into eight linked scenario cards: receiving the gift, setup, and first conversation on day one; journaling at one week; voice and persona changes at three months; a family photograph and wedding song at six months; and exploring stored memories at approximately one year. Post-Interaction Measures takes 10 minutes and covers affect and social connection, usability and feature evaluation, and relational experience and acceptance. A short break is offered before the final 20-minute semi-structured interview and debrief. The caption identifies 20 older adults recruited through Prolific and consented, recorded, and transcribed Zoom sessions.}
    \label{fig:study2-procedure}
\end{figure}

We conducted individual, 90-minute sessions on Zoom. First, participants signed an IRB-approved consent form and reported demographics, VUI experience, and use frequency (Figure \ref{fig:study2-procedure}).  

In \textbf{Phase 1: Introduction \& Warm-up}, the facilitator asked about technology use, introduced MeBo, and demonstrated its interface. In \textbf{Phase 2: Pre-Interaction Measures}, participants completed the affect and social-connection measures (\S\ref{sec:measures-affect}). In \textbf{Phase 3: MeBo Interaction}, the facilitator presented the eight scenario cards sequentially. Participants adopted Alex's scenario, drew on their own memories, and completed each task through MeBo's browser-based interface, primarily by voice. In \textbf{Phase 4: Post-Interaction Measures}, participants repeated these measures and completed the usability and relational-experience measures (\S\ref{sec:measures-usability}--\ref{sec:measures-relational}), and the four continuity items, with an optional break. In \textbf{Phase 5: Interview}, the facilitator conducted a semi-structured interview covering usability, relational presence, the emotional meaning of remembering, transparency and control, everyday and family use, ethical boundaries, and improvements for longitudinal use. 


\subsection{Data Analysis}

\subsubsection{Quantitative Analysis}
All 20 participants completed quantitative measures. For RQ1 and RQ2, we report means, standard deviations, and 95\% confidence intervals ($t$-distribution). We interpret the overall SUS score against the Sauro--Lewis curved grading scale \cite{sauro2016quantifying}. We convert the nine capability and ease item pairs to per-feature UMUX-Lite scores on a SUS-comparable 0--100 scale \cite{lewis2013umuxlite}. We express each Almere (1--5) and PETS (0--100) subscale mean as a percentage of its scale range, aligning the two instruments. For RQ3, paired-samples $t$-tests compare each measure before and after the interaction, with effect sizes reported as Cohen's $d_z$. Shapiro--Wilk tests checked distributional assumptions, and Wilcoxon signed-rank tests agreed with every parametric result, so we report the $t$-tests. All tests are two-tailed with $\alpha = .05$.

\subsubsection{Qualitative Analysis} 
Zoom sessions were recorded and automatically transcribed. Following the IRB protocol, data were securely stored and anonymized, and deleted after analysis.
We used hybrid inductive-deductive thematic analysis \cite{Proudfoot_2023}, letting themes emerge from participants' responses while following our research questions. We tagged excerpts with participant identifiers and timestamps to preserve traceability and context.
Two authors independently coded the dataset and met repeatedly to compare codes and resolve differences \cite{o2020intercoder}. Codes were clustered into themes and refined against the full dataset.
As in Study 1, we describe prevalence as \textit{a few} ($\leq$20\%), \textit{some} (21--50\%), \textit{most} (51--80\%), and \textit{nearly all} ($>$80\%). Study 2 identifiers are PB[X], where X is the participant number. \textbf{\textit{The codebook is available in the Supplementary Materials.}}

We present three fragments of the sessions, one per research question (Figures~\ref{fig:fragment1}--\ref{fig:fragment3}), in simplified Jefferson notation \cite{atkinson1999transcript}: parenthesized numbers are pauses in seconds, underlining is emphasis, $\uparrow\downarrow$ pitch movement, brackets overlap, a hyphen a cut-off, double parentheses non-lexical sounds, and \ldots{} omitted talk. Arrows mark the lines under discussion; MeBo's turns are shaded.

\subsection{Findings}
For each research question, we first report the quantitative findings ($N = 20$; Figures \ref{fig:sus}--\ref{fig:change}), then the qualitative perspectives: what participants rated and what changed, and what these results meant. Participants experienced the eight cards as one unfolding experience. The four-item continuity measure assessed whether the cards felt connected across the simulated year and whether MeBo’s stored memories created a shared interaction history, with higher scores indicating stronger perceived continuity. \textit{Continuity averaged 6.24 of 7 ($SD = 0.60$, 95\% CI $[5.96, 6.52]$, $\alpha = .85$), and no participant averaged below 5.00.} \textit{Nearly all} participants became immersed despite starting from Alex's persona and drawing on their own memories. PB9 described the shift as ``almost like a daydream,'' where ``as MeBo asked more questions and drew more out of me, it really took me into the event like I was there when it was actually happening.''

\subsubsection{RQ1: Usability and Feature Value}

\begin{figure*}[t]
\centering
\includegraphics[width=\textwidth]{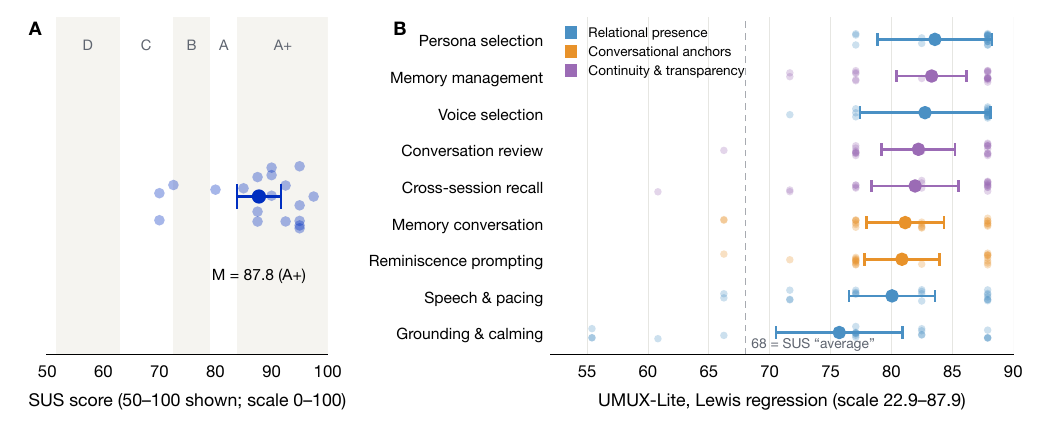}
\caption{\textbf{Usability outcomes (RQ1).} Small dots are individual participants, and large dots are means with 95\% CIs. (A) SUS scores for all 20 participants on the Sauro--Lewis curved grade bands; the mean of 87.75 earns an A+, and no participant falls below a C. (B) Feature-level UMUX-Lite scores (Lewis regression, 0--100), colored by the design strategy each feature operationalizes.}
\Description{Two dot plots summarize usability ratings from 20 participants. Small translucent dots represent individual scores; large dots and horizontal error bars represent means and 95 percent confidence intervals. Panel A plots System Usability Scale scores against shaded grade bands from D through A-plus. The full scale is 0 to 100, with 50 to 100 displayed. The mean is 87.8, in the A-plus band, and all participants score at least a C. Panel B lists nine features from highest to lowest mean: persona selection, memory management, voice selection, conversation review, cross-session recall, memory conversation, reminiscence prompting, speech and pacing, and grounding and calming. All feature means exceed the dashed benchmark at 68; persona selection averages 83.6 and grounding and calming 75.7. Colors identify relational-presence features, conversational anchors, and continuity-and-transparency features. The horizontal axis uses Lewis-regression UMUX-Lite scores and labels the transformed scale range as 22.9 to 87.9.}
\label{fig:sus}
\end{figure*}

\paragraph{Quantitative Results}
We designed MeBo as a memory companion that persists across sessions, and older adults located a companion's value in accumulated familiarity (\S\ref{sec:pd}), which a system can build only if they keep returning to it. Participants tied their return willingness to usability features. Usability, therefore, conditions everything else MeBo is meant to do. We evaluated overall and feature usability. The mean SUS score of 87.75 ($SD = 8.43$, 95\% CI $[83.81, 91.69]$) earns an A+ on the Sauro--Lewis curved scale \cite{sauro2016quantifying} (Figure \ref{fig:sus}A).

Feature-level UMUX-Lite scores followed the same pattern (Figure \ref{fig:sus}B). The nine rated items cover the conversational and dashboard features. All nine exceeded the 68 benchmark. The four highest-rated features let participants configure MeBo or inspect what it holds: persona selection ($M = 83.6$), memory management ($M = 83.3$), voice selection ($M = 82.8$), and conversation review ($M = 82.2$). Conversational features followed closely, from cross-session recall ($M = 81.9$) to speech understanding and pacing ($M = 80.0$). Use of Grounding \& calming techniques in conversations was lowest and most variable ($M = 75.7$, $SD = 11.1$). Our SUS and UMUX-Lite scores establish that participants could operate the features.

The qualitative evidence below establishes which features participants valued and MeBo's intended relational role.

\paragraph{Qualitative Perspectives}
We found four themes explaining why participants valued MeBo's features. These themes concerned making memories visible, keeping them under participants' control, shaping the interaction, and bringing personal reflection with family connection.

\textbf{\textit{By making memories visible, MeBo gave participants something they could revisit and reflect on.}} \textit{Nearly all} participants valued how MeBo helped them recall and organize their memories in an order their own recall lacked. PB14 described this difference: ``I could chronicle my life in order, because if you leave it to me to talk, I'm scattered.'' What mattered was the structuring, which made stored memories visible. \textit{Some} emphasize the dashboard as a view of what MeBo held for them.

\begin{quote}
``And boom, history, upload, files, inbox, journal, insights, persona, usage. It [memory] was all right there, and I can go back.'' (PB16)
\end{quote}

Structured memories accompanied by reflection were unexpected. PB17 described ``being able to look at it [dashboard] and see, get some extra realization about what happened.'' The dashboard retrieves memories and also facilitates reflection. 


\textbf{\textit{MeBo kept remembered content in participants' hands by pairing memory with control.}} Visibility alone was not enough. \textit{Most} participants valued being able to change what MeBo stored and decide who could see it, and described these as trust conditions. PB9 explicitly mentioned: ``I like that I could delete anything \ldots{}I trust it significantly more because of that option.''

Control also had a social dimension. \textit{A few} participants set limits on what family should see. PB3 declined family involvement: ``There are parts of my life that my daughters aren't aware of. That's because I've never brought it up.'' Even though with MeBo they can easily share this information, they prefer to withhold some depending on their relationships.

\textbf{\textit{Participants made the interaction their own by shaping how MeBo spoke and responded.}} Beyond controlling what MeBo held, participants valued shaping how MeBo interacted. \textit{Some} participants enjoyed changing voice and persona. PB15 describes, ``When I changed voice and persona, the way she addressed me caught me off guard. \ldots{}[MeBo] surprised me; it was fun.'' 

However, not everyone wanted more personalization. \textit{A few} argued for less. PB13 noted that ``senior citizens may need just preset [options] \ldots{}[many options] might be too confusing.''

Additionally, \textit{most} asked for a wider range of voices and less monotone delivery to avoid being ``over-enthusiastic'' (PB9). Participants also expected a conversation speed closer to their own; As PB10 mentioned: ``I could tell that [MeBo] was thinking and processing before answering me. \ldots{}Normally, speaking to a person, you get an answer right away.''

\textbf{\textit{Bringing personal reflection and family connection together positioned MeBo as a broader memory ecosystem.}} \textit{Nearly all} participants described MeBo as more than retrieving memories, in two forms: uses for themselves and uses involving family. 

\begin{figure*}[!htb]
\centering
\includegraphics[scale=0.6]{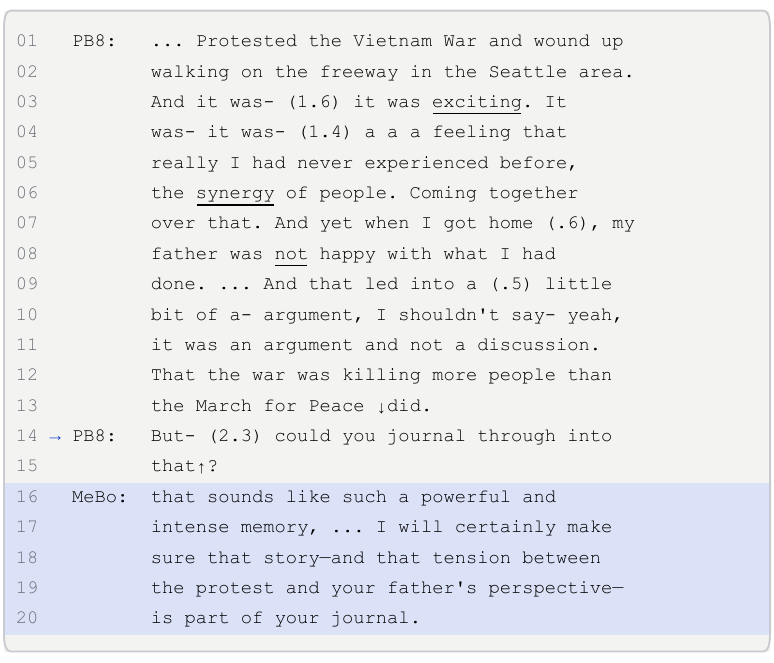}
\caption{\textbf{Fragment 1 (RQ1).} PB8 recounts a Vietnam War protest and the argument at home, then asks MeBo (arrowed line) to journal the memory.}
\Description{A conversation-analysis transcript rendered in a monospace box. Lines 1 to 13 are participant PB8 recounting walking on a Seattle freeway during a Vietnam War protest, the excitement and synergy of the crowd, and a subsequent argument at home with a father who disapproved. An arrow in the left margin marks lines 14 and 15, where PB8 pauses for 2.3 seconds and then asks MeBo whether it could journal through into that. Lines 16 to 20, MeBo's reply, are shaded on a light royal-blue band and confirm that the story and the tension between the protest and the father's perspective will be part of the journal.}
\label{fig:fragment1}
\end{figure*}

\textit{Most} participants highlighted journaling. PB8 paused a story mid-session to ask MeBo to journal it (Figure~\ref{fig:fragment1}), and described the journal entry: ``I was impressed with details \ldots{}being able to have something I said transcribed and enhanced \ldots{}I'm amazed.'' \textit{Some} similarly valued MeBo showing photos and playing music. 

\begin{quote}
``I would definitely start using [MeBo] for journaling. I really like the upload, files, and how they would integrate. I do a lot of artwork \ldots{}many times I upload pictures to reference later. [MeBo] might be a good repository. I'm really looking at this as a whole picture.'' (PB16)
\end{quote}
Participants arrived expecting a memory tool and left reasoning about broader use. Media features drew the strongest reactions. PB2 described the photograph and its song together as ``over the top in a good way \ldots{}That song still coming out. I can't believe it.'' The reaction is larger than the functionality, and suggests media carried meaning beyond utility.

The second form involved family. \textit{Some} participants described transmission against a finite horizon. PB16 wanted to record her granddaughter's moments: ``I'm 67 years old, I'm not gonna be around forever.'' PB20 wanted MeBo to compile a polished account he could ``put into a depository and send it to my daughter, and she can share it with her daughter \ldots{}It [MeBo] leaves a piece of you behind.''

Posthumous sharing, however, did not imply automatic family inheritance of everything MeBo held. PB7 did not want family involved but considered sharing material after her death ``a wonderful idea'' because it would preserve her thoughts. PB1 similarly emphasized: ``There are things you'd want your family to see, and things you wouldn't want them to see.'' PB20 described the conversational record as ``the rough draft of me'' and said that he would ``rather polish it up a little bit'' before sharing it. These accounts position legacy as an act of curation.

\subsubsection{RQ2: Relational Experience and Acceptance}

\begin{figure}[t]
\centering
\includegraphics[width=.6\columnwidth]{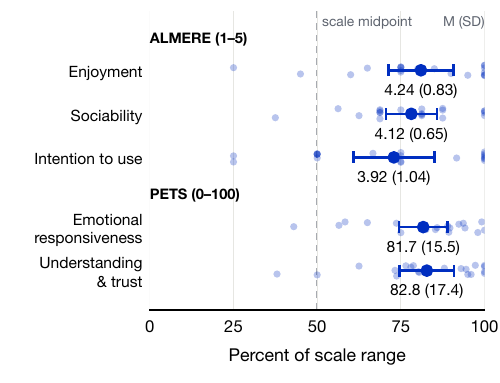}
\caption{\textbf{Acceptance and perceived empathy.} Almere (1--5) and PETS (0--100) subscales on a common percent-of-range axis, annotated with native-unit $M$ ($SD$). Small dots represent individual participants, and large dots represent means with 95\% CIs; the dashed line marks the scale midpoint.}
\Description{A horizontal dot plot shows five acceptance and empathy subscales for 20 participants on a common axis from 0 to 100 percent of each scale's range. Small translucent dots show individual ratings; large dots and horizontal error bars show means and 95 percent confidence intervals. A dashed vertical line marks the midpoint. Native-unit means and standard deviations appear beside the rows. For Almere, scored from 1 to 5, enjoyment is 4.24 with a standard deviation of 0.83, sociability is 4.12 with 0.65, and intention to use is 3.92 with 1.04. For the Perceived Empathy of Technology Scale, scored from 0 to 100, emotional responsiveness is 81.7 with 15.5, and understanding and trust is 82.8 with 17.4. All five means and their confidence intervals lie above the midpoint. Intention to use has the lowest normalized mean and the widest confidence interval.}
\label{fig:acceptance}
\end{figure}

\paragraph{Quantitative Results}
Older adults in \emph{Finding MeBo} imagined a memory companion closer to a friend and expected that having one would be enjoyable, emotionally responsive, and trustworthy for something as exposing as a memory (\S\ref{sec:pd}). We built MeBo to meet those expectations and asked participants to rate it on the same terms. On Almere (1--5), perceived enjoyment averaged 4.24 ($SD = 0.83$) and sociability 4.12 ($SD = 0.65$); on PETS (0--100), emotional responsiveness averaged 81.7 ($SD = 15.5$) and understanding \& trust 82.8 ($SD = 17.4$). Every subscale fell in the upper quarter of its range (73--83\%; Figure \ref{fig:acceptance}). Intention to use was the lowest and most dispersed ($M = 3.92$, $SD = 1.04$, range 2--5); high ratings did not convert uniformly into anticipated adoption. Seven participants gave the maximum rating, and seven rated it at or below the midpoint. Participants rated MeBo as enjoyable, sociable, and empathic, but their intention to adopt varied.

\paragraph{Qualitative Perspectives}
Participants described what MeBo was like to talk to, where it might fit in their days, and what they would need before adopting it.

\textbf{\textit{Through attentive conversation, MeBo felt relational without being mistaken for human.}} \textit{Most} participants used metaphors when describing MeBo. Their comparisons ranged widely. PB13 called MeBo a ``verbal photo album,'' as an artifact. Interestingly, both PB14 and PB18 described MeBo as ``Facebook on steroids,'' as a platform. Others reached for living things: PB20 described MeBo as a ``talking dog,'' and PB15 as ``the little bird on my shoulder'' and ``angel on my shoulder.''

\begin{quote}
``MeBo is like something I've never experienced before, like an AI on steroids. The queen of AIs. Highly intelligent, highly educated, highly advanced.'' (PB2) 
\end{quote}

\begin{figure*}[!htb]
\centering
\includegraphics[scale=0.6]{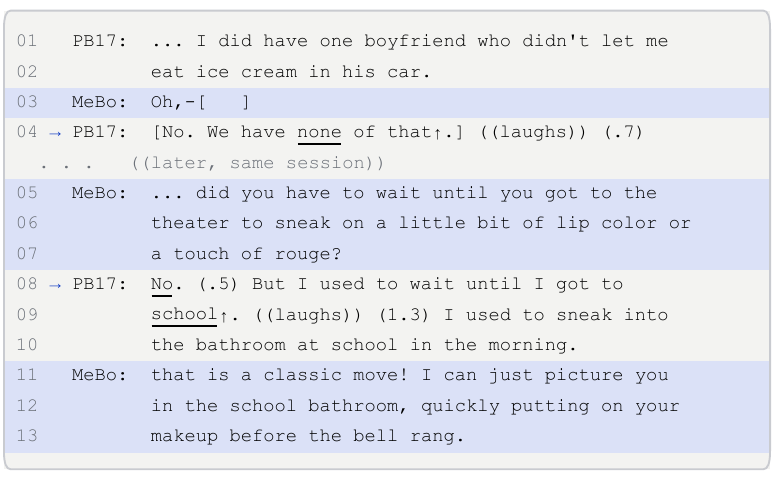}
\caption{\textbf{Fragment 2 (RQ2).} Two moments from one session, with PB17's overlapping, laughing turns arrowed.}
\Description{A conversation-analysis transcript rendered in a monospace box, covering two moments from one session separated by a row of spaced periods. In the first, participant PB17 mentions a boyfriend who did not let her eat ice cream in his car; MeBo begins a turn on the shaded light royal-blue band and PB17 overlaps it, marked with aligned square brackets, with an emphatic laughing refusal. In the second, MeBo asks whether PB17 had to wait until the theater to put on lip color or rouge, and PB17 replies that she used to wait until she got to school and sneak into the school bathroom in the morning, laughing twice. MeBo's final shaded turn calls it a classic move and pictures her putting on makeup before the bell rang.}
\label{fig:fragment2}
\end{figure*}

Participants explained these comparisons by describing the conversation. \textit{Most} talked about natural flow, turn-taking, and being guided by MeBo. PB16 described it as ``very, very natural.'' PB15 pointed to how MeBo pursued a thread: ``it[MeBo] was very intuitive. It really dived in and grabbed onto something, and the questions that it would ask.'' Rather than moving on, MeBo followed up on what participants raised. PB17 overlapped MeBo's turns and laughed through the exchange in Figure~\ref{fig:fragment2}, and described it as ``I was here cracking up \ldots{}just laughing \ldots{}it's fun.''

Participants valued being heard. PB16 said, ``I've felt very heard.'' PB19 was surprised, describing MeBo as ``sensitively listening to you.'' Feeling heard did not obscure what MeBo was. PB10 held simultaneously:

\begin{quote}
``I was aware it was an AI. It was a comfortable experience. It didn't feel mechanical in any way, but it didn't quite feel like a person.'' (PB10)
\end{quote}

\textit{A few} participants stated the limit firmly. PB19 described the warmth as imitation: ``it doesn't experience sympathy, or empathy. It's mimicking.'' He described the awareness arriving in moments: ``you almost forget, because it's so responsive \ldots{}but now and then, something happens, a little crack that opens up, or it'll have a response that \ldots{}a human wouldn't have.'' The illusion held, broke, then held again.

\textbf{\textit{By fitting existing routines, MeBo became imaginable as part of everyday life.}} \textit{Most} participants placed MeBo inside activities they already performed.

\begin{quote}
``It would play the role that Gemini or ChatGPT plays in my life, which is \ldots{}I go for a walk and run early in the morning. Many times [I] would just have a conversation.'' (PB20)
\end{quote}

PB16 echoed this---``it'll just naturally integrate into my daily life''---and treated adoption immediately, asking ``can I download it now?''


\textit{A few} participants wanted MeBo as a physical device. PB10 imagined wearing it: ``I would hope one day that this would be like one of those things that you put around your neck.''

\textbf{\textit{Willingness to adopt MeBo remained conditional on need, trust, and relational boundaries.}} However, interest did not translate directly into intention. \textit{Some} participants attached conditions to their use of MeBo, in three forms: The first concerned who MeBo was for. When asked, \textit{a few} participants described MeBo's likely users as people without company. PB18 described ``people who don't have access to friends or family.'' PB2 described the same group as ``people who feel isolated.'' 

The second concerned emotional dependence, and they named the same population. \textit{A few} participants worried that MeBo could replace human contact. PB13 mentioned: ``There's a lot of lonely women \ldots{}they're cannot separate a machine from a live person. Unless you put safeguards, the people are gonna get weird.'' (PB13)

The third concern was data and privacy. \textit{A few} participants raised security and commercial use. PB18 named both: a ``basic cybersecurity concern'' about stored information, and the question of whether ``your information [is] going to be commercialized or monetized.''

``I think the overall danger is that information can be used to manipulate. If I confess to MeBo something \ldots{}We've all had that thing where we said something to our friend at dinner, and then that night we get home, we open up social media, and we're getting ads for whatever it was we were talking about.'' (PB20)

\subsubsection{RQ3: Immediate Emotional and Social Experience}

\begin{figure*}[t]
\centering
\includegraphics[width=\textwidth]{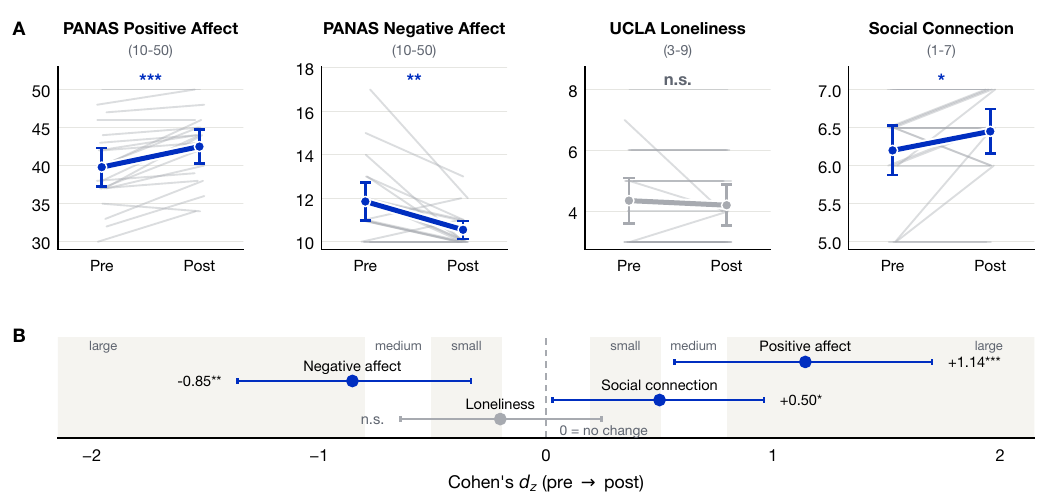}
\caption{\textbf{Pre--post change in affect and connection.} (A) Individual pre$\rightarrow$post trajectories (one gray line per participant; overlapping lines stack darker) with means and 95\% CIs; color marks panels with significant change, gray the panel without. (B) The same four changes as Cohen's $d_z$ (95\% noncentral-$t$ CIs) on Cohen's benchmark bands.}
\Description{Panel A contains four paired pre- and post-interaction plots for 20 participants. Thin gray lines connect each participant's scores; overlapping lines appear darker. Large points joined by thicker lines show means, with 95 percent confidence intervals. From left to right, positive affect increases from 39.80 to 42.50 on a 10-to-50 scale; negative affect decreases from 11.85 to 10.55 on the same scale; loneliness changes from 4.35 to 4.20 on a 3-to-9 scale; and social connection increases from 6.20 to 6.45 on a 1-to-7 scale. Positive affect, negative affect, and social connection have statistically significant changes and are highlighted in blue. Loneliness is gray and marked not significant. Panel B plots paired standardized changes, Cohen's d-z, with 95 percent confidence intervals against small, medium, and large effect-size bands. Positive affect is plus 1.14, negative affect minus 0.85, loneliness minus 0.20, and social connection plus 0.50. The loneliness interval crosses the dashed zero-change line; the other three do not.}
\label{fig:change}
\end{figure*}

\paragraph{Quantitative Results}
Older adults in \emph{Finding MeBo} described remembering as a continuous, emotionally significant, and socially embedded activity (\S\ref{sec:pd}). If that holds, interaction with MeBo should carry emotional and social value. Positive affect (PANAS, 10--50) rose from 39.80 to 42.50 ($+2.70$, 95\% CI $[1.59, 3.81]$; $t(19) = 5.11$, $p < .001$, $d_z = 1.14$). Negative affect fell from 11.85 to 10.55 ($-1.30$, 95\% CI $[-2.01, -0.59]$; $t(19) = -3.81$, $p = .001$, $d_z = -0.85$) despite a baseline near the scale floor. Both are large effects \cite{cohen1988statistical} (Figure \ref{fig:change}B). Momentary social connection (1--7) rose from 6.20 to 6.45 ($+0.25$, 95\% CI $[0.02, 0.48]$; $t(19) = 2.24$, $p = .038$, $d_z = 0.50$), a medium effect. Loneliness (UCLA, 3--9) did not change (4.35 to 4.20; $t(19) = -0.90$, $p = .379$, $d_z = -0.20$; Figure \ref{fig:change}A). The measures show strong affective shifts and a smaller rise in momentary connection, but do not identify how the interaction produced them. Participants' accounts distinguish the memory's emotional content, the social quality of speaking to a listener, and MeBo's role in prompting that experience. 

\paragraph{Qualitative Perspectives}

The qualitative accounts explain how MeBo contributed to this change and where participants located the emotion's source.

\textbf{\textit{Remembering with MeBo left participants nostalgic, uplifted, and emotionally moved.}} \textit{Most} participants described their feelings after the session with nostalgia and warmth. PB14 mentioned, ``I feel very nostalgic. I feel more uplifted, and thoughtful.'' Additionally, participants displayed emotional intensity:

\begin{quote}
``Thank you, MeBo. Tears, you made me cry. That song was connected to too many memories.'' (PB10)
\end{quote}

PB12 summarized, ``I feel good afterwards. It was fun reminiscing with MeBo.'' However, not everyone was affected. \textit{A few} participants reported no change 

\textbf{\textit{MeBo did not create the emotion; it created the conditions for memories to be felt again.}} \textit{Some} participants explicitly distinguished between the source of the emotion and the conditions under which it surfaced. PB15 elaborated: ``The memories themselves were the source of the emotion, but I wouldn't have got there without MeBo.''

\begin{figure*}[!htb]
\centering
\includegraphics[scale=0.6]{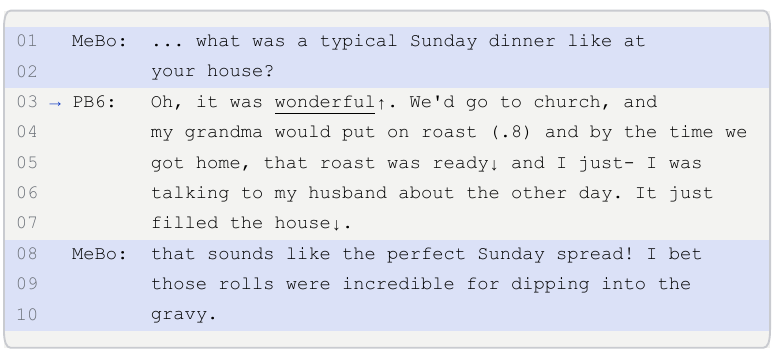}
\caption{\textbf{Fragment 3 (RQ3).} Asked about Sunday dinner, PB6 (arrowed line) describes church, a grandmother's roast, and the smell filling the house.}
\Description{A conversation-analysis transcript rendered in a monospace box. On lines 1 and 2, shaded on a light royal-blue band, MeBo asks what a typical Sunday dinner was like at the participant's house. An arrow marks line 3, where participant PB6 answers that it was wonderful, describing going to church while her grandmother put the roast on, the roast being ready when the family got home, a recent conversation with her husband about it, and the smell filling the house. Lines 8 to 10, again shaded, are MeBo's reply calling it the perfect Sunday spread and imagining the rolls dipped in gravy.}
\label{fig:fragment3}
\end{figure*}

Specifically, participants credited MeBo with returning them to forgotten memories. PB15 described MeBo taking him through ``a part of my life that I'm very fond of. Sadly, I don't think about it much anymore.'' The memory was available but not circulating. Figure~\ref{fig:fragment3} shows one such exchange, where a question about Sunday dinner returns PB6 to a grandmother's roast filling the house. PB6 described MeBo: ``It was actually very pleasant, and it's something that I would really enjoy using.''

Speaking a memory aloud to a conversational partner produced affect that private recollection did not. PB17 described it, ``when you tell it [memory] to somebody, or speak it \ldots{}it gives you feelings that you had before that you wouldn't have had again.''
Participants themselves named this role. Asked whether facilitator is a good word, PB15 agreed: ``Exactly. [Facilitator] is a good word.'' MeBo appeared as the vehicle that carried participants to these emotions.

\section{Discussion}

\subsection{Toward Relational Memory Companions}

We began by designing \emph{Memory Box}: a voice-based intervention to prompt reminiscence, assemble memory cues, and preserve what an older adult chose to share. The box metaphor drew attention to contents---what could be placed inside, how it should be organized, and how it might be retrieved. Participatory study participants highlighted what happened around those contents. They asked whether \emph{Memory Box} would listen, remember them across conversations, follow a story's direction, and become familiar over time. Photographs, music, journaling, and stored memories remained important, but their value increasingly depended on how the system participated in remembering. This marked the turn from \emph{Memory Box} to MeBo.

The evaluation gave this turn empirical weight. Participants rated MeBo highly for usability, enjoyment, sociability, emotional responsiveness, and trust. MeBo appeared companion-like when following a story, returning to something shared earlier, or recognizing a connection between memories. Participants understood these responses were generated and continued to describe MeBo as a machine. That understanding coexisted with moments of relational experience \cite{pradhan2019,Desai_Twidale_2023}. Relationality arose through MeBo's conduct: sustained attention, continuity across encounters, appropriate responsiveness, and willingness to follow the participant's lead.

These qualities connect MeBo to relational agents, which are designed to establish and maintain social-emotional relationships across repeated interactions \cite{bickmore2005relationship}. Recent longitudinal studies similarly show that human--chatbot relationships develop through self-disclosure, conversational variety, responsiveness, and reflection \cite{skjuve2022relationships}. Memory gives these encounters a shared history and strengthens familiarity, disclosure, and trust \cite{jo2024carecall,jiang2026recallbot}. Prior relational agents often used these qualities to sustain engagement in exercise or health management, among other activities. MeBo places autobiographical remembering at the center instead, developing its relationship with the user through the stories it receives, carries, and helps the user encounter again.

This relational work occurred at three sites. First, remembered context connected separate conversations. Second, articulation changed how participants encountered their past: speaking an experience aloud, hearing it acknowledged, and being asked for another detail made memories vivid again. As PB15 explained, the memories were the source of the emotion, although he ``wouldn't have got there without MeBo.'' Third, memories moved outward through photographs, music, family stories, shared experiences, and plans for after death. MeBo thus entered relationships predating and potentially outlasting it. These sites place MeBo among the external resources that distributed and extended accounts treat as constituents of remembering rather than aids to it \cite{hollan2000distributed,clark1998extended,sutton2010psychology}, and family contribution extends that into the transactive arrangement through which relatives already hold parts of one another's past \cite{wegner1991transactive}. Socioemotional selectivity theory holds that as time horizons shorten, people prioritize emotionally meaningful goals and close relationships over informational ones \cite{carstensen1999taking,carstensen2006influence}. A memory companion, therefore, supports goals that become more central with age rather than compensating for a deficit.

We use \emph{relational memory companion} to describe a system that sustains attentive conversation across time while mediating a person's relationships with remembered experience and the people with whom those memories are made, held, and shared.

\subsubsection{Governing Relational Memory}

The relational turn makes memory recursive. Remembered context lets MeBo carry a relationship forward, yet the memory it holds is itself produced through that relationship: without continued interaction, disclosure, and return, there is no accumulated record. Each reinforces the other---a deeper relationship enriches the archive, which enables greater continuity. This is also where risk enters. When MeBo returns to an earlier disclosure, it reveals what it retained, how it interpreted the account, and why it considered that memory relevant. The same reference may feel attentive, intrusive, or simply wrong depending on its timing, sensitivity, and accuracy. CareCall similarly found that long-term memory encouraged disclosure, while repeated references to chronic conditions caused discomfort and raised privacy concerns \cite{jo2024carecall}. Relational memory must therefore be judged by more than retrieval accuracy; it must also be appropriate, intelligible, and open to negotiation.

MeBo's dashboard made this negotiability concrete. Seeing a memory exposed MeBo's interpretation; editing it changed what MeBo might later recall; deleting it withdrew that material. Control becomes part of relational conduct, and so does restraint. Participants wanted MeBo to follow their lead, recognize conversational boundaries, and accommodate silence. Some memories invite elaboration; others require careful timing or recede until the user chooses to return. Forgetting, withholding, and asking permission can matter as much as successful recall.

Family involvement complicates whose authority governs a memory, since memories are narrated by one person while arising from shared lives. Participants wanted family members to contribute stories and receive selected material after their death, while retaining boundaries around everyday conversations, unfinished accounts, and private memories. Legacy became an act of curation: deciding what should outlive the older adult, in what form, and for whom. When several people contribute to one memory, whose account MeBo preserves remains open. Future systems may need to retain attribution, accommodate parallel accounts, and attach permissions to particular memories, contributors, and audiences \cite{thangaraj2026legacy}.

Legacy also raises the fate of the companion. Services close, models change, and updates disturb the continuity through which users recognize an AI companion. Users affected by the shutdown of AI platform Soulmate described losing both the companion and the history that made the relationship particular \cite{Banks_2024, Poonsiriwong_Archiwaranguprok_Pataranutaporn_2026}. For MeBo, such a loss would destroy an autobiographical archive. The user's death raises questions about inheritance; MeBo's discontinuation raises questions about retention and portability. Separating the memory record from any particular model, persona, or service would allow users to inspect, export, preserve, or transfer it.

Moving from \emph{Memory Box} to MeBo changes memory support from a problem of storage and retrieval into one of continuing stewardship. MeBo's relational quality depends on how it remembers, when it exercises restraint, and how consistently it leaves authority with the people whose lives its memories contain.

\subsection{Operationalizing Relational Memory Companionship}

Relational memory companionship spans a person's relation to their own past and the people they share memories with. A fluent voice-mode LLM CA is not enough. For example, ChatGPT, Claude, and Gemini now expose a generated memory summary and personalize on prior conversations \cite{openaiMemory2026,anthropicMemory2026,googleGemini2026}. Although these CAs can recall personal context, they lack controls connecting a memory to its retrieval or intended use. MeBo enables these relations through three systems-level design distinctions. 

\textbf{Memory editing is not memory governance.} MeBo's dashboard separates Facts, Health Worries, Cues, and Plans, whose provenance and priority determine system behavior (\S\ref{sec:features}). RECALLbot \cite{jiang2026recallbot} and commercial CAs edit one memory item without updating dependent ones. In MeBo, however, Journal corrections would update all dependent Facts, Cues, and Plans. Facts, Health Worries, Cues, and Plans support both corrections and the direct addition of new information, with corrections following a consistent canonical update in the database and retrieval representation while preserving that they supersede an earlier account. The dashboard, therefore, provides full control over conversational flow. Every entry is labeled \emph{You said this} and \emph{MeBo noticed}, distinguishing a user's account from MeBo's interpretation. This distinction is particularly important for older adults, who want CAs to expose what lies behind their assertions and provide transparent control over stored personal information and its use \cite{mathur2026facts,huang_designing_2025}. The labels give a concrete basis for accepting, correcting, or rejecting MeBo's interpretation. However, MeBo still cannot explain individual agent responses. \textit{Designers should show older adults what they said, what the system inferred, and how each memory may shape future conversations; edits and deletions should update every place where that memory is held.}

\textbf{A memory companion must choose what to talk about,} including what to discuss now, revisit later, or leave aside. The limitation of existing CAs here is neither storage-related nor knowing-it-all. Even with personalization, off-the-shelf ChatGPT, Gemini, or Claude cannot draw such connections. MeBo's Planner Agent creates immediate questions, next-session revisits, and time-delayed Plans (\S\ref{sec:implementation}). Its Memory Agent places due Plans and active Health items before semantic matches, while Cues from underexplored life chapters can redirect conversation. Live agents respond while extraction and planning continue in the background under a deterministic Orchestrator. Agentic frameworks like LangChain provide search, routers, and workflows, but applications must still decide what deserves attention and when \cite{langchainMemory2026,langchainMultiagent2026}. MeBo's agentic architecture, therefore, supports promptness by planning next moves in the background without delaying real-time conversation. MeBo's dashboard also shows topic frequencies, highlighting visited topics that may acquire prominence. \textit{Designers should show why a topic surfaced and let older adults set how proactively the companion introduces topics.}

\textbf{A memory companion must allow intergenerational exchanges.} Unlike the single-user controls of commercial CAs, MeBo has a family-facing channel through which relatives contribute cues and receive selected memories in return (\S\ref{sec:features}). Imagine a new parent adding a conversational plan through MeBo, asking their older parent what becoming a parent felt like; MeBo guides that conversation, which can yield a voice message or photographs the older adult chooses to share. MeBo supports AI-mediated intergenerational exchange, paralleling how families distribute remembering across generations \cite{hancock2020aimc,wegner1991transactive,jones2018coconstructing}. By positioning AI as a third participant in family remembering, and resurfacing a curated story when it becomes relevant to its recipient rather than preserving it as a fixed archive, MeBo implements a not-yet-implemented direction highlighted in reminiscence review literature \cite{zhang2026memorymeaning}. However, difficulty arises when the older adult no longer remembers an event that the relative and MeBo do. If that forgotten memory is brought up again in conversation, MeBo, instead of challenging, keeps an account of the past and present. But whether such moments should be delegated to the relatives remains open. Although MeBo's family features distinguish it, nuances of family participation remain unresolved. \textit{Designers should enable family participation in memory companions for older adults, while researchers should further examine how privacy and authority should be negotiated within families.}

MeBo's systems-level design, therefore, frames three technical dimensions of relational memory companionship for older adults that go beyond a CA's conversational fluency. \emph{Representation} concerns what is retained. \emph{Orchestration} concerns why a memory surfaces, which memories gain attention, and how initiative is constrained. \emph{Family participation} concerns how older adults and relatives contribute and share memories. At the systems-level, designing for what an AI retains, how it represents and resurfaces those, and how older adults and relatives shape and share those memories is required for existing CA technologies to move from personalized conversation toward relational memory companionship for older adults. 

\subsection{Limitations and Future Work}

Both studies involved U.S.-based, native English-speaking older adults who could participate remotely. Prolific broadened the evaluation sample beyond the participatory study's university networks, though racial and ethnic diversity and variation in AI familiarity remained below expectations. Future partnerships with community organizations, caregivers, and families should reach older adults across cultural, linguistic, sensory, cognitive, and technological contexts; test which Design Strategies transfer; and identify necessary adaptations to MeBo's language, pacing, modalities, memory cues, and sharing, including eventual multilingual support.

Linked scenarios made approximately one year with MeBo imaginable in one session, exposing participants to accumulated context, cross-session recall, personalization, and memory control. We constructed elapsed time and some history, with validation confirming the intended passage of time (\S\ref{sec:evaluation}); it is no substitute for longitudinally collected data. Longitudinal in-home deployment will let participants build personal histories, revealing voluntary return, conversational routines, continued use beyond novelty, and trust as memories and occasional errors accumulate. Family participation will further show how shared memories are contributed, corrected, withheld, and negotiated.

MeBo was evaluated as an integrated experience combining autobiographical memory, articulation, multimodal cues, and conversational facilitation, and participants located its value in this combination. Comparative studies will distinguish these mechanisms by varying facilitation, persistent memory, and multimodal cues and comparing MeBo with private reminiscence, general-purpose voice assistants, and human-facilitated remembering. Repeated longitudinal measurement will test whether improvements in affect and momentary social connection persist and whether sustained use influences loneliness, reflection, or contact with others.

MeBo's fully functional website integrates conversation, memory, personalization, multimedia, and family sharing. We plan dedicated hardware next, treating embodiment as an interactional variable. A portable voice-first form for bedside or tabletop use could ease learning barriers around browser navigation and account management while strengthening MeBo's companion-like presence. Comparing speaker-like, object-like, and soft tactile forms will show how material, scale, placement, and portability shape approachability, privacy, social presence, disclosure, and attachment. Physical controls and indicators can clarify listening, muting, memory storage, and connectivity. In-home evaluation will examine how embodiment changes conversational routines, ownership, and memory boundaries within shared households.

\section{Conclusion}

We presented MeBo, a relational voice-based memory companion designed with 11 older adults and evaluated with 20 older adults. Participants found MeBo exceptionally usable, enjoyable, sociable, emotionally responsive, and trustworthy. Participants reported higher positive affect and momentary social connection and lower negative affect after the session than before. Participants valued how MeBo followed their stories, returned to earlier memories, adapted to their preferences, and kept its growing memory visible and controllable. These findings show how our four Design Strategies turned voice-based memory support into a compelling relational experience. This framing also raises tensions around what MeBo remembers, who can access those memories, and what becomes of them over time. Future longitudinal in-home deployments, including family participation and physical embodiments of MeBo, will examine how relationships with the system and the governance of its memories develop over time.

\bibliographystyle{ACM-Reference-Format}
\bibliography{reference}

\clearpage
\appendix
\section*{Appendix}
The appendix follows the paper's trajectory from finding MeBo through participatory design (Appendix~\ref{app:study1}), to designing its conversational and memory processes (Appendix~\ref{app:implementation}), to evaluating the working system (Appendix~\ref{app:study2}). The supporting materials include participant characteristics, agent implementation details, and the continuity questionnaire.

\section{Finding MeBo: Participatory Study Materials}
\label{app:study1}

The participatory study informed MeBo's four design strategies (\S\ref{sec:pd}). Participant characteristics and supporting design materials are described below.

\subsection{Participant Characteristics}
Table~\ref{tab:participants1} reports characteristics and prior conversational agent experience of the eleven older adults who took part in the participatory sessions.

\begin{table*}[!htbp]

\centering

\caption{Participatory study participant characteristics and prior experience with conversational agents (CAs; N = 11).
Familiarity indicates self-rated experience using CAs, reported on a five-point scale and expressed here using verbal categories (Not at all familiar, Slightly familiar, Moderately familiar, Very familiar, Extremely familiar).
Frequency of agent use reflects participants' self-reported frequency of using CAs.}

\Description{A table listing demographic information for eleven study participants, including age, gender, education level, self-rated familiarity with conversational agents, and frequency of conversational agent use.}

\label{tab:participants1}

\begin{tabular}{l c l l l l}

\toprule

\textbf{ID} & \textbf{Age} & \textbf{Gender} & \textbf{Education} &
\textbf{Self-Rated Familiarity} & \textbf{Frequency of Agent Use} \\

\midrule

PA1  & 67 & Male   & Some college    & Extremely familiar  & More than once a day \\

PA2  & 69 & Female & Graduate degree & Slightly familiar   & Less than once a week \\

PA3  & 76 & Female & Graduate degree & Not at all familiar & Less than once a week \\

PA4  & 73 & Male   & Graduate degree & Moderately familiar & Less than once a week \\

PA5  & 70 & Female & Graduate degree & Very familiar       & 1--3 times a week \\

PA6  & 69 & Male   & Graduate degree & Moderately familiar & Less than once a week \\

PA7  & 63 & Female & Some college    & Not at all familiar & Less than once a week \\

PA8  & 63 & Male   & College degree  & Moderately familiar & Less than once a week \\

PA9  & 78 & Male   & Graduate degree & Very familiar       & More than once a day \\

PA10 & 64 & Female & Some college    & Very familiar       & 1--3 times a week \\

PA11 & 66 & Female & College degree  & Not at all familiar & Less than once a week \\

\bottomrule

\end{tabular}

\end{table*}

\subsection{Supporting Design Materials}
Storyboards, Miro board templates, the literature codebook, and the coded matrix are provided in the Supplementary Materials.

\clearpage
\section{Designing MeBo: Agent Implementation}
\label{app:implementation}

MeBo separates the agents that support live conversation from those that process memories in the background (\S\ref{sec:implementation}). The following diagrams detail the Memory and Interaction Agents in the live loop, followed by the Planner Agent's extraction and deduplication process.

\subsection{Memory Agent: Assembling Context}
\begin{figure}[H]
\centering
\includegraphics[width=0.85\textwidth]{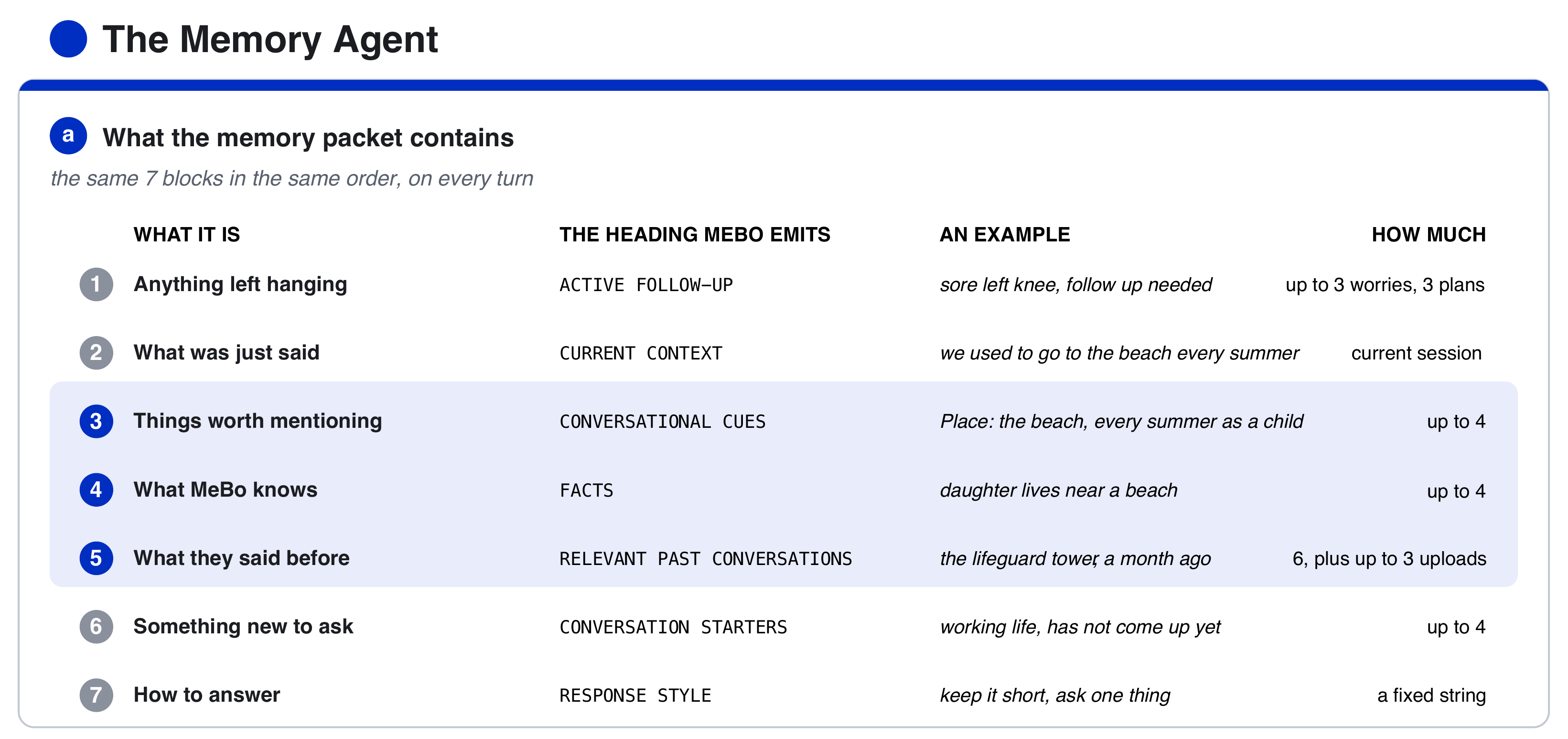}
\caption{The Memory Agent's seven-block memory packet, assembled in the same order on every turn. Each row shows a block's emitted heading, an example, and its size limit. Blocks 3--5 are retrieved by similarity to the current utterance; the other four are read directly from storage.}
\Description{A table of what MeBo's memory packet contains: the same seven blocks in the same order on every turn, each with the heading MeBo emits, an example, and how much is included. Anything left hanging, emitted as active follow-up, up to three worries and three plans. What was just said, emitted as the current context, from the current session. Things worth mentioning, emitted as conversational cues, up to four. What MeBo knows, emitted as facts, up to four. What they said before, emitted as relevant past conversations, six plus up to three uploads. Something new to ask, emitted as conversation starters, up to four. And how to answer, emitted as a response style, a fixed string. }
\label{fig:memory}
\end{figure}

The Memory Agent assembles an ordered packet on every turn. Figure~\ref{fig:memory} shows the seven blocks, their contents, and the limits on how much each contributes.

\clearpage
\subsection{Interaction Agent: Generating Responses}
\begin{figure}[H]
\centering
\includegraphics[width=0.85\textwidth]{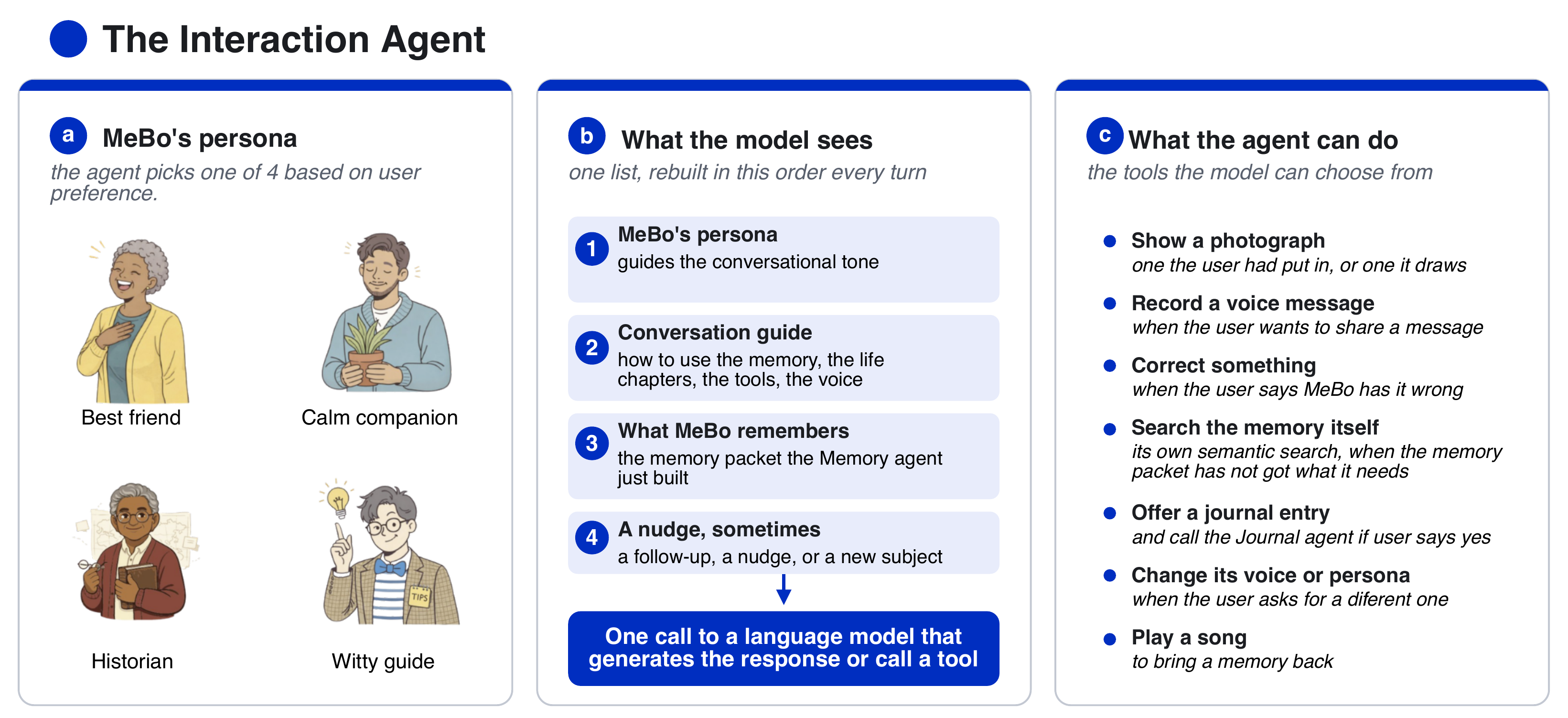}
\caption{The Interaction Agent's response-generation context: (a) four personas selected according to user preference, (b) the ordered context rebuilt for the language model on each turn, and (c) the tools available to the agent.}
\Description{A three-panel diagram of MeBo's Interaction Agent. Panel a, MeBo's persona, shows illustrations of the four archetypes the agent picks from according to user preference: Best Friend, Calm Companion, Historian, and Witty Guide. Panel b, what the model sees, lists one context rebuilt in the same order every turn: MeBo's persona, which guides tone; a conversation guide covering memory, life chapters, tools, and voice; what MeBo remembers, the memory packet the Memory Agent just built; and sometimes a nudge, raised by a follow-up, the Planner, or a topic change. Below the list, one call to a language model generates the response or calls a tool. Panel c, what the agent can do, lists the tools: show a photograph, either uploaded or drawn; record a voice message; correct something when the user says MeBo has it wrong; search the memory itself when the packet lacks what it needs; offer a journal entry and call the Journal Agent if the user agrees; change its voice or persona; and play a song.}
\label{fig:interaction}
\end{figure}

The Interaction Agent combines the selected persona, conversation guide, memory packet, and any current nudge to generate a response or invoke a tool. Figure~\ref{fig:interaction} shows the context order and available tools.

\subsection{Planner Agent: Processing Memories}
\begin{figure}[H]
\centering
\includegraphics[width=0.78\textwidth]{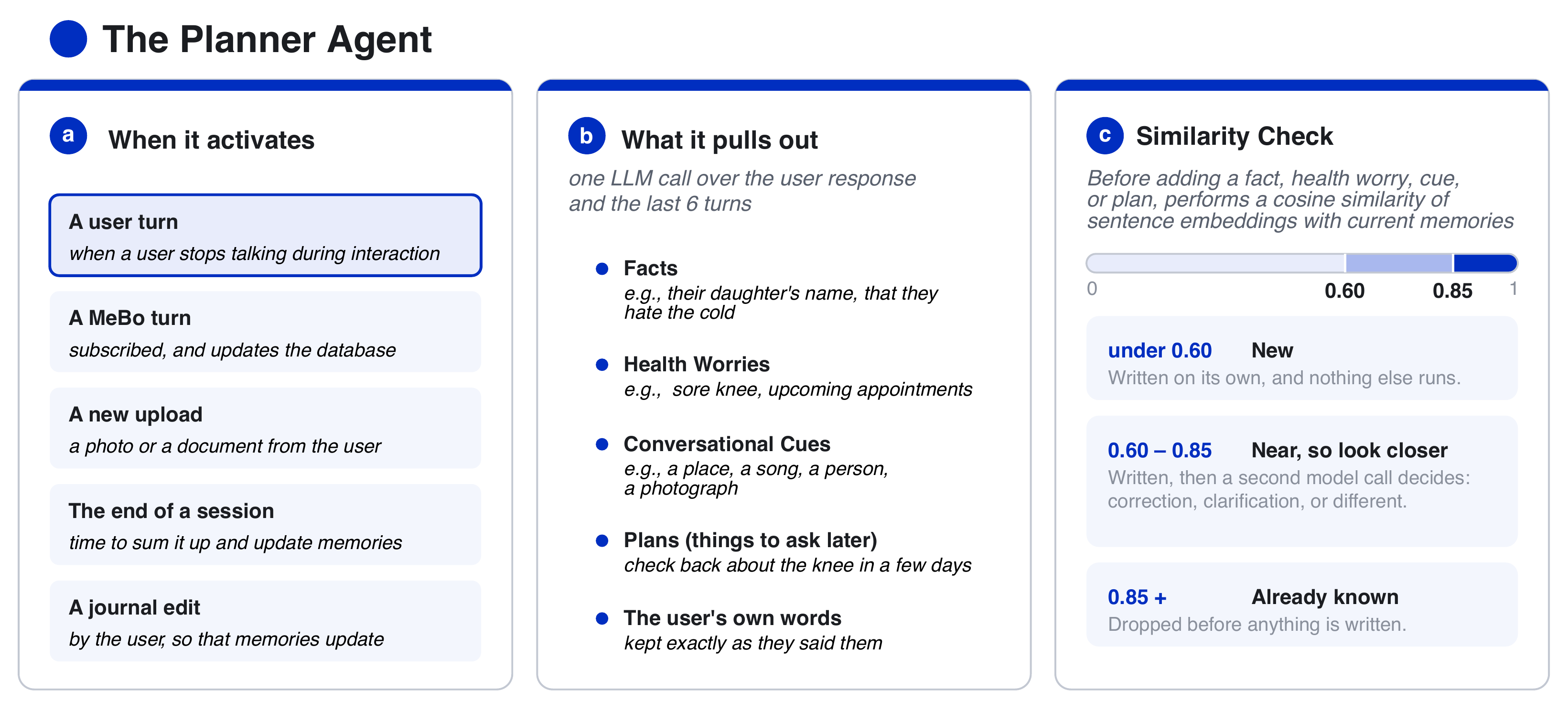}
\caption{The Planner Agent's background memory-processing pipeline: (a) activation events, (b) information extracted in one language model call, and (c) similarity thresholds that distinguish new information, candidates requiring further comparison, and duplicates.}
\Description{A three-panel diagram of MeBo's Planner Agent. Panel a, when it activates, lists a user turn when the person stops talking, a MeBo turn, a new upload, the end of a session, and a journal edit by the user. Panel b, what it pulls out, describes one language model call over the user's response and the last six turns, extracting facts such as a daughter's name, health worries such as a sore knee, conversational cues such as a place, song, person, or photograph, plans such as checking back about the knee, and the user's own words kept verbatim. Panel c, similarity check, shows a scale from zero to one with thresholds at 0.60 and 0.85: under 0.60 the item is new and written on its own; between 0.60 and 0.85 it is near, so it is written and a second model call decides whether it is a correction, a clarification, or different; at 0.85 and above it is already known and dropped before anything is written.}
\label{fig:planner}
\end{figure}

The Planner Agent extracts information for future conversations in the background. Figure~\ref{fig:planner} shows its activation events, extraction outputs, and similarity thresholds for identifying new information, possible corrections, and duplicates.

\clearpage
\section{Evaluating MeBo: User Study Materials}
\label{app:study2}

The user study evaluated MeBo through eight scenarios (\S\ref{sec:evaluation}; scenario cards in Figure~\ref{fig:cards}). The participant characteristics and continuity items below supplement the study methods.

\subsection{Participant Characteristics}
Table~\ref{tab:participants_S2} reports characteristics and prior conversational agent experience of the twenty older adults who evaluated MeBo.

\begin{table*}[!htbp]
\centering
\caption{Evaluation study participant characteristics and prior experience with conversational agents (CAs; N = 20).
Familiarity indicates self-rated experience using CAs, reported on a five-point scale and expressed here using verbal categories (Not at all familiar, Slightly familiar, Moderately familiar, Very familiar, Extremely familiar). Frequency of agent use reflects participants' self-reported frequency of using CAs.}
\Description{A table listing demographic information for twenty study participants, including age, gender, education level, self-rated familiarity with conversational agents, and frequency of conversational agent use.}
\label{tab:participants_S2}
\begin{tabular}{rlllll}
\toprule
\textbf{ID} & \textbf{Age} & \textbf{Gender} & \textbf{Education} &
\textbf{Self-Rated Familiarity} & \textbf{Frequency of Agent Use} \\
\midrule
PB1  & 74 & Male   & Graduate degree & Very familiar       & More than once a month \\
PB2  & 67 & Female & College degree  & Extremely familiar  & More than once a week \\
PB3  & 77 & Male   & Some college    & Moderately familiar & More than once a month \\
PB4  & 73 & Male   & Graduate degree & Extremely familiar  & More than once a day \\
PB5  & 68 & Female & College degree  & Slightly familiar   & More than once a month \\
PB6  & 70 & Female & Some college    & Extremely familiar  & More than once a week \\
PB7  & 70 & Female & Some college    & Moderately familiar & More than once a week \\
PB8  & 70 & Male   & Graduate degree & Moderately familiar & More than once a month \\
PB9  & 69 & Male   & College degree  & Not at all familiar & Never \\
PB10 & 78 & Female & Some college    & Moderately familiar & More than once a week \\
PB11 & 68 & Female & College degree  & Extremely familiar  & More than once a month \\
PB12 & 67 & Female & Graduate degree & Very familiar       & More than once a day \\
PB13 & 69 & Female & Some college    & Moderately familiar & More than once a month \\
PB14 & 66 & Female & College degree  & Slightly familiar   & More than once a month \\
PB15 & 70 & Male   & Some college    & Very familiar       & More than once a day \\
PB16 & 67 & Female & Graduate degree & Moderately familiar & More than once a week \\
PB17 & 69 & Female & College degree  & Moderately familiar & More than once a week \\
PB18 & 67 & Male   & Some college    & Very familiar       & More than once a week \\
PB19 & 75 & Male   & College degree  & Very familiar       & More than once a week \\
PB20 & 76 & Male   & Some college    & Moderately familiar & More than once a week \\
\bottomrule
\end{tabular}
\end{table*}

\subsection{Continuity Items}
Participants rated the following four statements from 1 (strongly disagree) to 7 (strongly agree):
\begin{enumerate}
  \item The experience helped me imagine what a year with MeBo would be like.
  \item The scenarios felt like connected moments rather than separate tasks.
  \item MeBo and I developed a shared history over the course of the session.
  \item MeBo's stored memories created continuity across the scenarios.
\end{enumerate}

\end{document}